\documentclass[sigconf]{acmart}

\AtBeginDocument{%
  \providecommand\BibTeX{{%
    \normalfont B\kern-0.5em{\scshape i\kern-0.25em b}\kern-0.8em\TeX}}}
\usepackage{xspace}
\usepackage{amsthm}
\usepackage{amsmath}
\usepackage{enumitem}
\usepackage{booktabs}
\usepackage{multirow}
\usepackage{graphicx}
\usepackage[normalem]{ulem}
\useunder{\uline}{\ul}{}
\usepackage{float}
\usepackage{caption}
\usepackage{subcaption}
\usepackage{balance}

\newtheorem{definition}{Definition}[section]

\copyrightyear{2022}
\acmYear{2022}
\setcopyright{acmlicensed}\acmConference[WWW '22]{Proceedings of the ACM Web Conference 2022}{April 25--29, 2022}{Virtual Event, Lyon, France}
\acmBooktitle{Proceedings of the ACM Web Conference 2022 (WWW '22), April 25--29, 2022, Virtual Event, Lyon, France}
\acmPrice{15.00}
\acmDOI{10.1145/3485447.3512273}
\acmISBN{978-1-4503-9096-5/22/04}

\begin{document}

\title{Large-scale Personalized Video Game Recommendation via Social-aware Contextualized Graph Neural Network}

\author{Liangwei Yang}
\email{lyang84@uic.edu}
\affiliation{%
  \institution{University of Illinois at Chicago}
  \country{USA}
}

\author{Zhiwei Liu}
\email{zhiweiliu@salesforce.com}
\affiliation{%
  \institution{Salesforce Research}
  \country{USA}
}

\author{Yu Wang}
\email{ywang617@uic.edu}
\affiliation{%
  \institution{University of Illinois at Chicago}
  \country{USA}
}

\author{Chen Wang}
\email{cwang266@uic.edu}
\affiliation{%
  \institution{University of Illinois at Chicago}
  \country{USA}
}

\author{Ziwei Fan}
\email{zfan20@uic.edu}
\affiliation{%
  \institution{University of Illinois at Chicago}
  \country{USA}
}

\author{Philip S. Yu}
\email{psyu@cs.uic.edu}
\affiliation{%
  \institution{University of Illinois at Chicago}
  \country{USA}
}

\renewcommand{\shortauthors}{Liangwei et al.}

\begin{abstract}
Because of the large number of online games available nowadays, online game recommender systems are necessary for users and online game platforms. The former can discover more potential online games of their interests, and the latter can attract users to dwell longer in the platform. This paper investigates the characteristics of user behaviors with respect to the online games on the Steam platform. Based on the observations, we argue that a satisfying recommender system for online games is able to characterize: personalization, game contextualization and social connection. However, simultaneously solving all is rather challenging for game recommendation. Firstly, personalization for game recommendation requires the incorporation of the dwelling time of engaged games, which are ignored in existing methods.
Secondly, game contextualization should reflect the complex and high-order properties of those relations. Last but not least, it is problematic to use social connections directly for game recommendations due to the massive noise within social connections. To this end, we propose a Social-aware Contextualized Graph Neural Recommender System~(SCGRec), which harnesses three perspectives to improve game recommendation. We conduct a comprehensive analysis of users' online game behaviors, which motivates the necessity of handling those three characteristics in the online game recommendation.
\end{abstract}




\maketitle

\section{Introduction}
Recent years have witnessed the rapid growth of the video game industry. According to statistics\footnote{\url{https://www.statista.com/statistics/292056/video-game-market-value-worldwide/}}, the global video game market value in 2020 is $\$155.89$ billion and is expected to reach $\$268.81$ billion in 2025. 
Along with the rising of the market, values are massive new releasing video games. It is reported that on Steam platform\footnote{\url{https://store.steampowered.com/}}, $10,263$ new games are released in 2020 while the number was $276$ in 2010\footnote{\url{https://www.statista.com/statistics/552623/number-games-released-steam/}}. 
Due to the time limitation of users in playing video games, a game recommender system is necessary to assist users to discover new games of their potential interests.
In this way, we can improve users' experience in-game engagement and can thus attract players to stay longer within the platform.

Though designing game recommender systems attracts recent research attentions~\cite{de2021design,anwar2017game,bertens2018machine}, they are still under-explored.
A satisfying game recommender system should be able to characterize the following three perspectives: 1) \textbf{personalization}, modelling the game engagement interests of users, 2) \textbf{game contextualization}, revealing complex relations among games and 3) \textbf{social connection}, interpreting the game engagement of users as a type of social activities.
Though these factors have already been investigated in other recommendation scenarios~\cite{liu2020basconv,liu2020basket, wang2021dskreg, DBLP:conf/cikm/FanL00Y21, zheng2019gated, DBLP:conf/cikm/FanLZX0Y21, 10.1145/3404835.3463036, fan2022sequential}, directly adopting existing methods from other domains is problematic due to the unique challenges in tackling game recommendations. 

Firstly, the personalization of users should reveal their interests in games, which requires the awareness of game engagements. A comprehensive game engagements should incorporate: 1) \textit{which games} users have engaged in and 2) \textit{how long} users dwell in those games~\cite{yi2014beyond}. 
The former is widely tackled as the collaborative signals~\cite{cheuque2019recommender,he2020lightgcn} between users and games, while the latter is seldom studied. On one hand, dwelling time reflects the loyalty of users to games, which should be of high vitality in revealing the users' preference. On the other hand, because of the design of games (e.g., RPG v.s. MOBA), the dwelling time of different games may not be comparable~\cite{sifa2014playtime}, which also leads to the difficulty. 

Moreover, game contextualization is to comprehend the relatedness among games. Contextually similar games are recommended to users based on their historical game engagements.
Existing methods~\cite{cheuque2019recommender,de2021design} construct the context of games by formatting that side information as features, \textit{e.g.,} the developer and price of a game will be processed as one-hot vector and float number features, respectively.
This enables the recommender system to harness the side information of games rather than solely relying on the collaborative signals.
Nevertheless, digesting context as features is unable to characterize complex and high-order relations between games.
For example, if game A has the same developer as game B, and game B is usually co-purchased with game C according to users transactions, game A and game C should also be highly related when conducting recommendations. 
However, the complex co-purchase relation and the transitivity of relations are both neglected by feature-based methods. Hence, we should explicitly tackle the relations between games for contextualization. 

Last but not least, social connections are rather crucial in game recommendations because many online video game engagements are not isolated individual behaviors but social activities. 
In this sense, discovering potential interests of users from their friends enables the recommender system to identify the game engagements of users from a social perspective. 
Nevertheless, the challenge is that due to social inconsistency~\cite{yang21consisrec}, users' friends may be of different impacts concerning distinct games.
For example, users usually play MOBA games \textit{Dota 2}\footnote{\url{https://store.steampowered.com/app/570/Dota_2/}} with friends in a team, whereas single-player RPG games \textit{Portal}\footnote{\url{https://store.steampowered.com/app/400/Portal/}} are played individually.
As a result, directly recommending those games equally based on the engagements of friends~\cite{perez2020hybrid} yields sub-optimal performance. Therefore, we should infer the impacts of friends on users. 

To this end, we propose a novel game recommendation framework named SCGRec. SCGRec can be split into three parts, time-aware context aggregation, context-aware social aggregation and personalized predictor. User's contextual embedding is learned in time-aware context aggregation module. It aggregates information from user's engaged games with consideration on dwelling time. User's social embedding is obtained from context-aware social aggregation module. This module firstly learns attention weight of neighbors from user's contextual embedding, and obtains social embedding by weighted sum of neighbors' personalized embedding. In personalized predictor, contextual embedding and social embedding are fused with user's personalized embedding by weighted sum. The prediction score is obtained by a dot product of aggregated user embedding and personalized item embedding. The code and data for experiment is available online at \textcolor{blue}{\url{https://github.com/YangLiangwei/game-recommendation}}.

The contributions of this paper are as follows:

\begin{itemize}[leftmargin=*]
    \item We conduct comprehensive data analyses of the game engagement based on a large-scale Steam dataset containing nearly billions of interactions. We investigate both the statistical characteristics and underlying patterns of the game engagements.
    \item We propose a unified framework to incorporate personalization, game contextualization and social connections, which improves the game recommendation performance compared with existing methods, especially those state-of-the-art graph neural networks.
    \item We conduct extensive ablation studies to demonstrate the effectiveness of leveraging all those three perspectives. 
\end{itemize}

\section{Related Work}
\subsection{Game Recommendation}


Designing game recommender systems is not a academic focus until recent years~\cite{de2021design,anwar2017game,bertens2018machine} due to the rapid developments of game industries~\cite{marchand2013value}. We categorize existing works as: 1) the collaborative filtering~(CF)-based methods, 2) the content-based methods, and 3) the hybrid methods. 

Firstly, collaborative filtering~(CF)-based methods~\cite{DBLP:journals/advai/SuK09,zhou2021pure} harness the interactions between users and games and assume users with similar behaviors are of similar game preference.
R. Sifa etc.~\cite{sifa2014archetypal,sifa2014playtime}, are pioneering researchers solving game recommendation problems via archetypal, which factorizes the interaction matrix into the linear combination of archetypes. 
GAMBIT~\cite{anwar2017game} investigates both the item-based and user-based CF methods for game recommendation. G. Cheuque etc.~\cite{cheuque2019recommender} demonstrate the game recommendation performance on Steam with a simple CF-based method via the ALS algorithm~\cite{takacs2012alternating}. CF-based methods are effective in reflecting the interactive signals between users and games.

Secondly, content-based methods~\cite{pazzani2007content} leverage the profile of users and the description of games and predict their interaction likelihood. 
B. Paul etc.~\cite{bertens2018machine} predict the next game that users will interact by fitting their time-series patterns.
J. Kim etc.~\cite{kim2020sequential} conduct the ranking-based recommendation by sequentially considering user historical behaviors. 

Finally, hybrid methods characterize both the CF signals and content-based information for the recommendation. 
G. Cheuque etc.~\cite{cheuque2019recommender} investigate the performance of FM~\cite{rendle2012factorization} and DeepFM~\cite{DBLP:conf/ijcai/GuoTYLH17} towards game recommendation by leveraging both the interactions and feature information of users and games.
J. Pérez-Marcos etc.~\cite{perez2020hybrid} model feature-based relations between games and then apply the CF for the recommendation. They also firstly study the possibility of using social networks over users to improve performance.
Hybrid methods are of both flexibility and adaptability regarding the game recommendation task. Our proposed model is also a hybrid recommender system.

\subsection{Graph-based Recommendation}
We review some graph-based recommendation methods as we adopt the graph neural network~(GNN)~\cite{DBLP:journals/tnn/WuPCLZY21,DBLP:journals/corr/abs-2106-12484,wang2021explicit, DBLP:conf/bigdataconf/WangDCCLY21} for the game recommendation. Graph-based recommendation methods model user-item interactions as a bipartite graph~\cite{he2020lightgcn, berg2017graph, ying2018graph, shen2021powerful, mao2021ultragcn,DBLP:conf/sdm/ZhangM20}, with potential extensions to the heterogeneous graph with additional user-user social graph~\cite{yang21consisrec,DBLP:journals/algorithms/MensahGY20} and item knowledge graph~\cite{wang2019kgat, wang2019knowledge}. Graph-based methods mostly adopt GNN for learning nodes~(users and items) embeddings. GNNs aggregate information from neighbors and can incorporate first-order and even higher-order information if multiple layers are stacked. Such unique characteristics allow GNNs to effectively capture high-order collaborative signals, which are crucial signals for learning user and item embeddings in recommendation~\cite{he2020lightgcn, wang2019neural}.

Heterogeneous GNNs~\cite{liu2020basconv, wang2019kgat} are applied for recommendation for the combination of more side information such as knowledge graph and social network. Knowledge-aware GNNs recommendation~\cite{wang2019kgat} increasingly attracts attentions, intending to improve recommendation by connecting items with knowledge entities by various relationships.
KGAT~\cite{wang2019kgat} introduces a unified collaborative knowledge graph from the user-item interaction graph and knowledge graph and proposes a GNN for learning both recommendation and knowledge tasks.  KGNN-LS~\cite{wang2019knowledge} introduces label smoothing regularization on knowledge graphs to further enhance the item embeddings learning.

In addition to knowledge graph, GNNs also enables the modelling of user-user social graph as side information in recommendation. GraphRec~\cite{fan2019graph} learns graph attention network to assign different attention weights to different social neighbors. DiffNet~\cite{wu2019neural} models information diffusion process~\cite{DBLP:journals/corr/abs-1803-08378} in social graph to enlarge user's influence scope~\cite{DBLP:journals/entropy/GuoYCCGM20}. DANSER~\cite{wu2019dual} performs dual graph attention networks on social and user-item interaction networks separately and fuses the learned embedding by a policy-based fusion layer. ConsisRec~\cite{yang21consisrec} dynamically samples informative neighbors and performs aggregation considering different relation types.
FeSoG~\cite{liu2021federated} proposes a GNN-based social recommendation system under graph federated learning setting~\cite{he2021fedgraphnn}.
Our work considers the social information as the social context via attention mechanism.

Our work aims to enhance game recommendation with additional contextual information. Different from previous works, SCGRec is flexible enough to integrate both time-aware contextual information and social connection information. 
Instead of using explicit message passing as existing works, SCGRec models those side information as contextual embeddings.
As such, we can improve the recommendation performance by leveraging both game engagement information and additional contextual information.


\section{Data Analysis}
In this section, we first present an overview of the data statistics, then dive deep into the data characteristics by visualizing data distributions from personalization, game contextualization, and social connections perspectives. 

\subsection{Overview of Data Statistics}


\begin{table}[]
\centering
\caption{Number of different nodes in Raw Data}
\label{tab:raw_data_stat}
\begin{tabular}{ccccc}
\toprule
users & games & genres & engagements & social connections \\
\hline 
3.8M & 4,079 & 22 & 384.3M & 392.7M\\
\bottomrule
\end{tabular}
\end{table}

The raw data\footnote{\url{https://steam.internet.byu.edu/}} is crawled by Mark et al.~\cite{DBLP:conf/imc/ONeillVWZ16} through Steam Web API\footnote{\url{https://developer.valvesoftware.com/wiki/Steam_Web_API}}. Players and games form a highly unbalanced bipartite graph. Millions of players can play each game while each player only plays a small number of games, which makes the aggregated information highly unbalanced on the two sides. The following analysis is based on all the collected data.

\subsection{Personalization}\label{sec:Personalization}
\begin{figure}[ht]
     \centering
     \begin{subfigure}[b]{0.23\textwidth}
         \centering
         \includegraphics[width=1\textwidth]{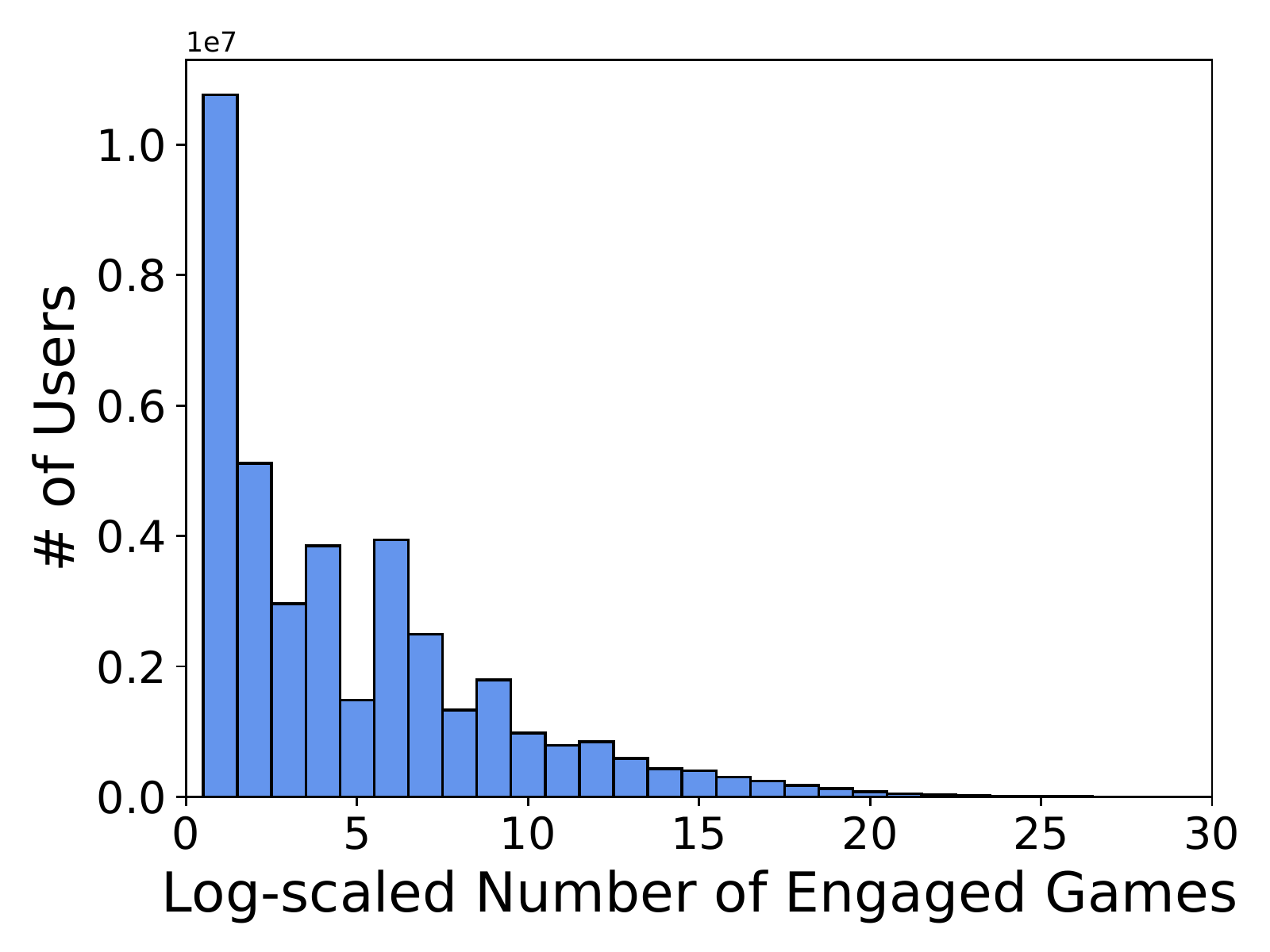}
         \caption{\# of Engaged Games}
         \label{fig:user_num_engages_games_dist}
     \end{subfigure}
     \hspace{-2mm}
     \begin{subfigure}[b]{0.23\textwidth}
         \centering
         \includegraphics[width=1\textwidth]{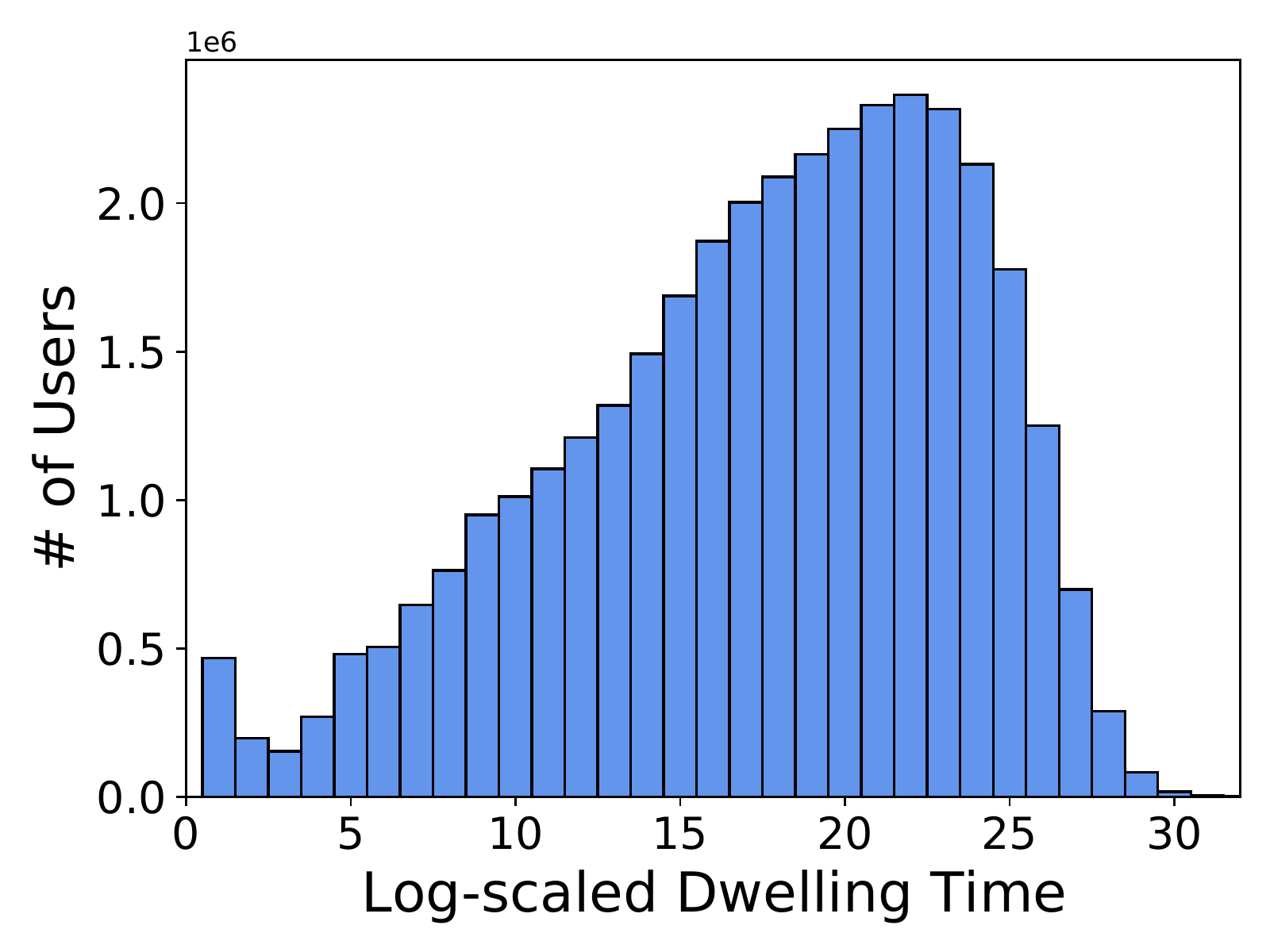}
         \caption{Dwelling Time}
         \label{fig:user_game_dwelling_time_dist}
     \end{subfigure}
     \caption{User Game Engagement Distributions}
\end{figure}

User engagement activities uncover their personalized interests in online video game platforms. We demonstrate the necessity and challenges of introducing personalization for game recommendation by investigating distributions of user engagements in both the numbers of engaging games and dwelling time. 
We visualize the user distributions with respect to the number of engaging games in Figure~\ref{fig:user_num_engages_games_dist} and dwelling time in Figure~\ref{fig:user_game_dwelling_time_dist}. 
From Figure~\ref{fig:user_num_engages_games_dist}, we can observe the long-tail distribution on the number of engaging games. We also show the number of users with respect to different number of interacted games in Table~\ref{tab:user_num_interaction}. It shows a large number of users only interact with limited number of games. Note that, users' personalized information is not restricted to historical game interactions. Users' social friends and dowelling time of each game all count as personalized information. SCGRec achieves personalized recommendation by considering all such information. In addition to the number of engaging games, we plot the user count distribution over different dwelling times in Figure~\ref{fig:user_game_dwelling_time_dist}.
Dwelling time uncovers how much time users engage in games and is also a crucial signal of user interests.
To illustrate this, we visualize the user count distribution on log-scaled dwelling time in Figure~\ref{fig:user_game_dwelling_time_dist}. The dwelling time distribution almost follows the normal distribution, which is different from the long-tail distribution of the number of engaging games. It reveals different behavioral patterns on whether and how long to engage in a game, which indicates that predictive modelings on these two signals demand distinct mechanisms. 


\subsection{Game Contextualization}
Game contextualization is to comprehend the relations between games, which can further interpret the recommendation~\cite{xu2020product}.
We investigate the game contextualization from both the genre level and the game level. Genre-level contextualization is to analyze whether games from different genres are co-played by users. 
\begin{figure}
         \centering
         \includegraphics[width=0.4\textwidth]{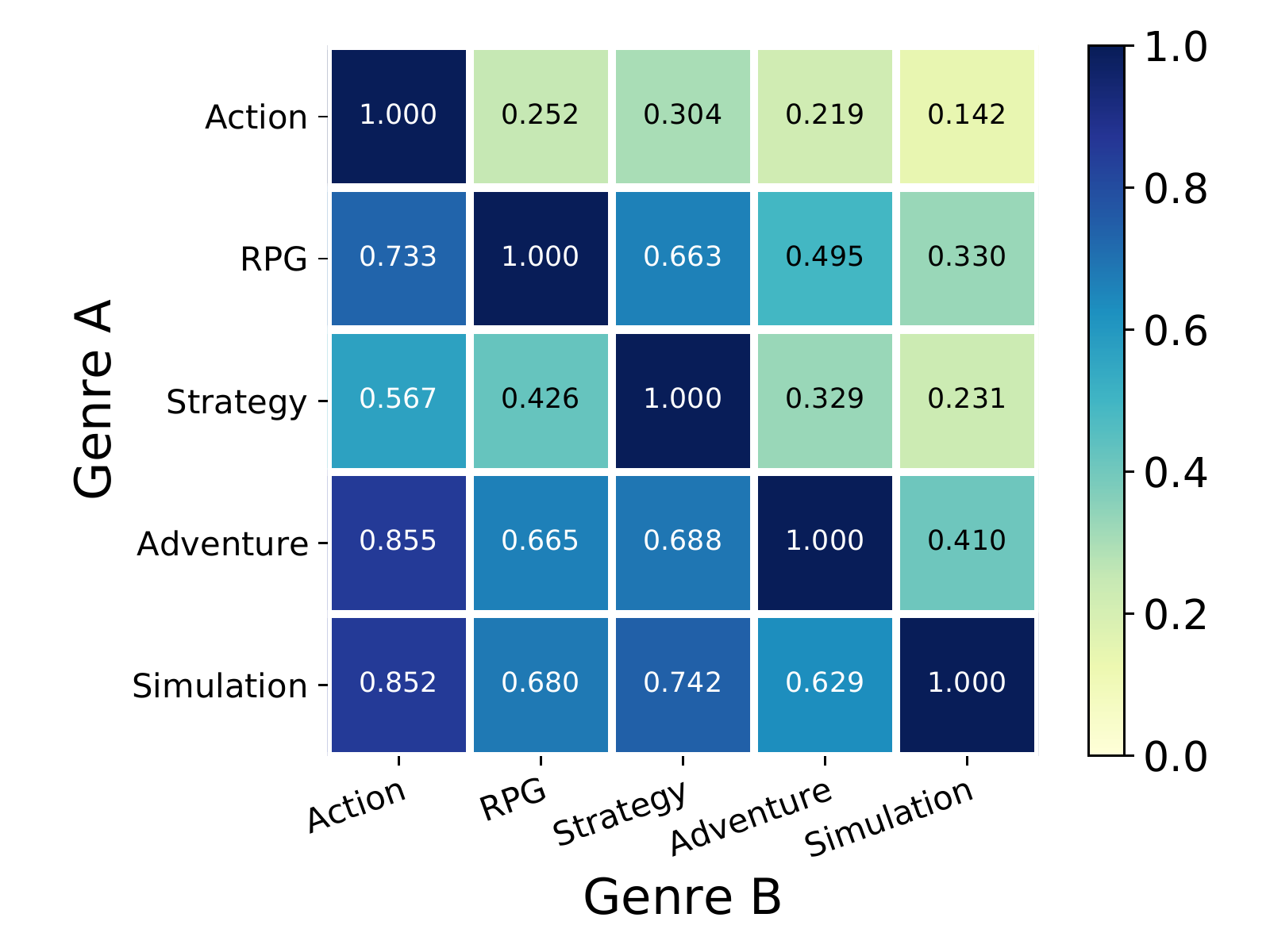}
         \caption{Genre Conditional Probability $P(A|B)$ Visualization}
         \label{fig:genre-genre condition}
\end{figure}
We select the five most popular genres. We calculate the probability that conditioned users engaged in one genre, how likely those users also play those other genres.  
The relations between genres are illustrated in Figure~\ref{fig:genre-genre condition}.
We observe that the conditional probability from different genres are diverse, which indicates the correlations between genres are distinct. 
Moreover, we observe that users playing simulation games are reluctant to play other game types because of the low probability in other genres.

\begin{figure}
         \centering
         \includegraphics[width=0.48\textwidth]{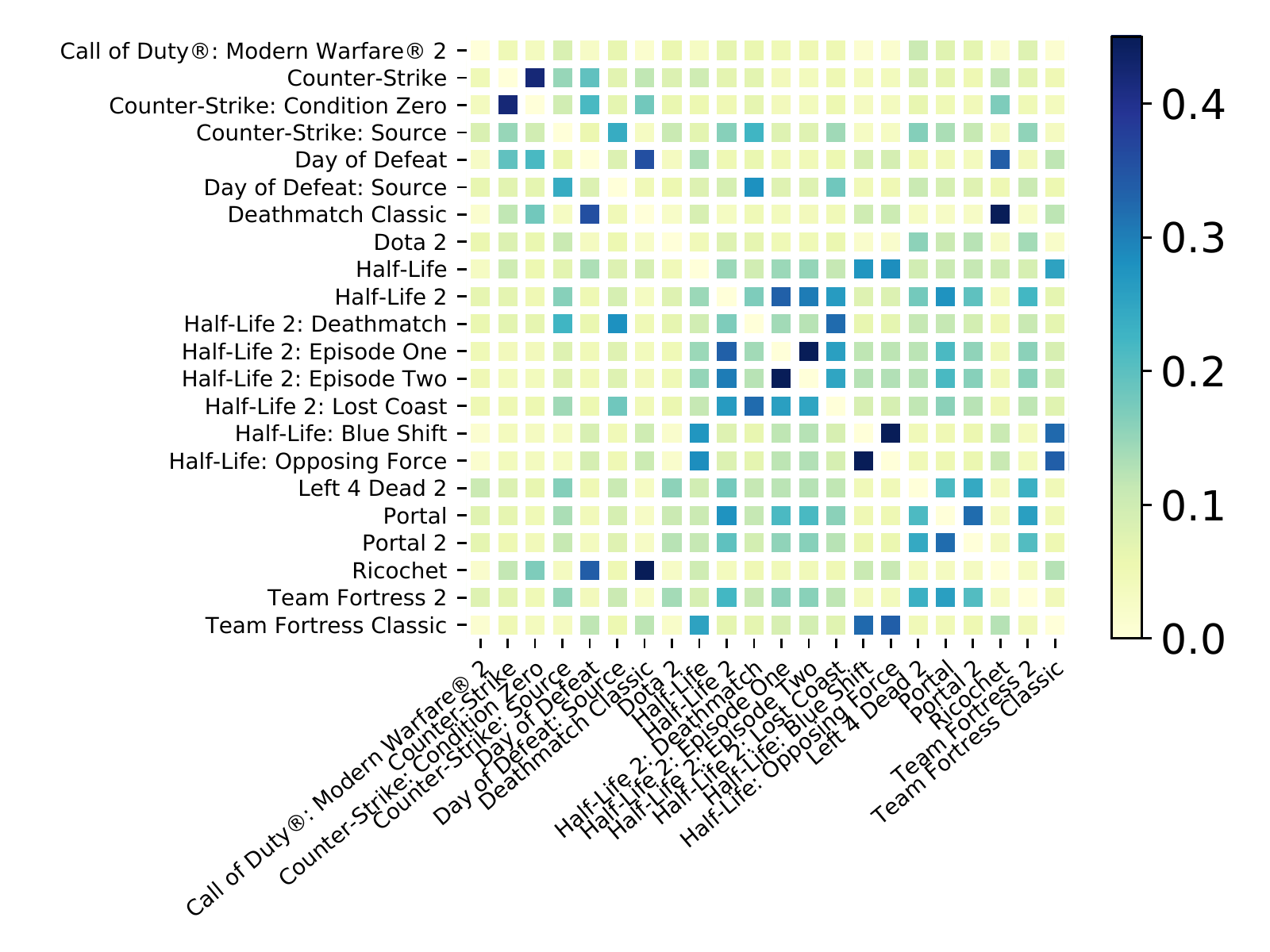}
         \caption{Game-Game Co-Purchase}
         \label{fig:Co-Purchase_relation}
    
\end{figure}

The second investigation is designed for game-level contextualization. Here, we present the co-purchase relation between games. 
Precisely, the co-purchasing relation score $c_{ij}$ between game $i$ and $j$ is calculated as follows:
\begin{equation}\label{eq:co_purchase}
{c_{ij}}=\frac{\text{\# of users engaged in both $i$ and $j$}}{\text{\# of users engaged in $i$} + \text{\# of users engaged in $j$}}.
\end{equation}
High relation scores represent that two games are played by the larger group of shared users, which implies their high relatedness.
We present the results for part of the games in Figure~\ref{fig:Co-Purchase_relation}.
We can observe that some game pairs have a high correlation, \textit{e.g.} the Counter-Strike and Counter-Strike: Condition Zero.
This will be a strong signal for game contextualization.
We also present the co-dwelling relation scores between those games in the appendix.

\subsection{Social Connections}\label{sec:social_connection}
Social activities also influence online video game engagements. To verify this, we conduct two types of data analyses over the game engagements of users and their friends. The first analysis is to demonstrate whether game engagements are social activities. 
To achieve this, we select the top-10 popular games and calculate the \textbf{Pearson correlation coefficient} between the dwelling time of users and the average dwelling time of their friends. 
The results are in Table~\ref{tab:Pearson_Correlation}. 
For comparison, we randomly select the same number of users with no friendships with the associated users and calculate their Pearson correlation coefficient. 
Observations are twofold. Firstly, game engagements are influenced by social connections because the correlation scores between friends are higher than between non-friends on all the games. 
Secondly, the intensity of social impacts is different across games. For example, the MOBA game \textit{Dota 2} has a higher correlation score compared with the single-player game \textit{Portal}. 
We also include the Pearson correlation coefficient concerning top-5 genres, and distribution of friend number in the appendix. 

\begin{table}[htbp]
\caption{Pearson correlation coefficient w.r.t. games.}
\label{tab:Pearson_Correlation}
\begin{tabular}{lcc}
\toprule
\textbf{Game}&\textbf{ user-friend} & \textbf{user-random}\\
\midrule
Counter-Strike & 0.5816 & 0.1846\\
Dota 2 & 0.5422 & 0.1649\\
Team Fortress Classic & 0.4705 & 0.0678\\
Team Fortress 2 & 0.4659 & 0.1489\\
Day of Defeat & 0.4325 & 0.0609\\
Left 4 Dead 2 & 0.4304 & 0.0917\\
Ricochet & 0.3084 & 0.0246\\
Deathmatch Classic & 0.2376 & 0.0142\\
Half-Life & 0.2264 & 0.0217\\
Portal& 0.1216 & 0.0254\\
\bottomrule
\end{tabular} 
\end{table}

\begin{figure}
 \centering
 \includegraphics[width=0.4\textwidth]{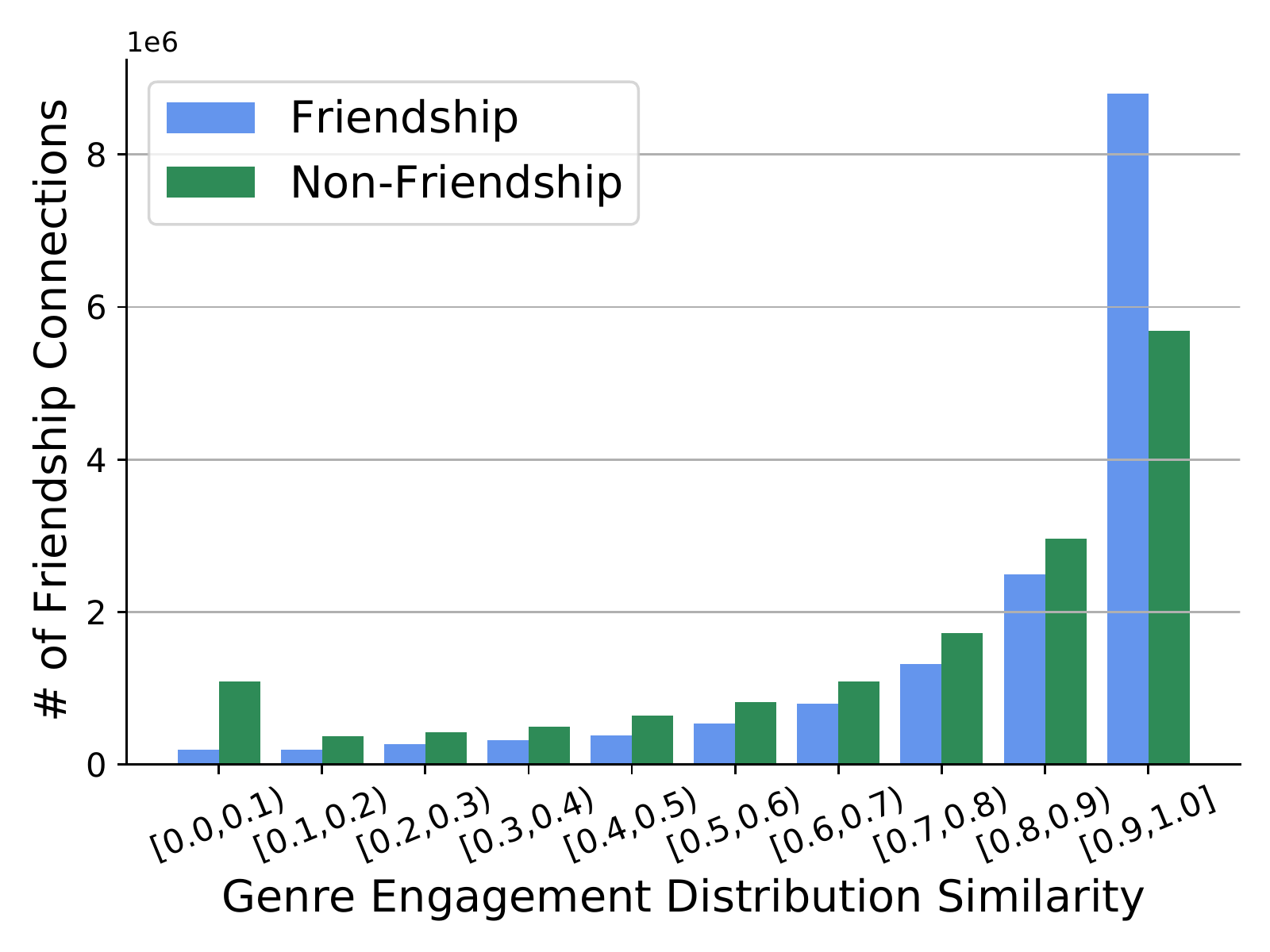}
 \caption{The comparison between counting distributions of friendship edges and non-friendship edges within different intervals of game genres engagement cosine similarity.}
 \label{fig:friendship_genre_cosine}
    
\end{figure}

The second analysis is to investigate whether friends are of similar game preference. To calculate this similarity score, we first define the user game preference vector as the genre-wise engagement distribution $[t_1, t_2, \dots, t_k]$, where each entry $t_k$ denotes the log-scaled dwelling time in genre $k$. Due to the power-law distribution of users in genres, we choose the top-5 genres.
Then, the preference similarity is calculated as the cosine similarity between users, computed over all social connections.
Furthermore, for comparison, we also randomly sample the same number of non-friendship links between users. The histogram of the similarity score is distributed in Figure~\ref{fig:friendship_genre_cosine}. 
We can observe that in the range of $[0.9,1.0]$, the number of friendship connections is more than non-friendship connections, suggesting that friends are more likely to have a similar preference. 

\section{Preliminaries}
In this section, we present several basic definitions before formulating the problem of game recommendation. 
\begin{definition}
\textbf{(Game Engagement)}. Given the game set $\mathcal{I}$, the game engagement of a user $u$ is defined as a set of (game,time) pairs as $\mathcal{E}_u = \{(i_1, t_1), (i_2, t_2), \dots, (i_k, t_k)\}$, where $i_1, i_2, \dots, i_k \in \mathcal{I}$ and $t_1,t_2, \dots, t_k \geq 0$. $k$ is the total number of engaged games of $u$ and $t_k$ denotes the dwelling time in game $i_k$.
\end{definition}

The personalization requires the game engagements of all users $\mathcal{U}$, which forms a bipartite attributed graph. The nodes are users and games, and the edges are their interactions with dwelling time. 
Dwelling time is calculated as how long the user has played a game, reflecting the intensity of the game engagement. 

Besides the interactions between users and games, we also leverage the context information for games, \textit{e.g.}, the developer of games. Moreover, instead of directly processing context information, we form relations between games via shared context.
In other words, if two games share the same context, we construct a link between them. In this way, we define the game context graph as follows:
\begin{definition}
\textbf{(Game Context Graph)}. Given the game set $\mathcal{I}$, the game context graph for context $c$ is defined as $\mathcal{G}^{(c)}_{i}=\{\mathcal{I}, \mathcal{E}_c\}$, where $\mathcal{I}$ is the node set and $\mathcal{E}_c$ denotes edges.
For two games $i,j\in\mathcal{I}$, $ (i,j) \in \mathcal{E}_c$ if they share context $c$. 
\end{definition}
The definition of context is rather flexible. Context shows some relevance between games. It can be the game attributes, co-engaged relations and game mechanisms~\cite{machado2019pitako}. Each relevance can be represented as one kind of edge in the game context graph. Those contexts constitute a multiple relation game context graph as $\mathcal{G}_{i}=\{\mathcal{G}_i^{(c)}\}|_{c=1}^{C}$, where $C$ is the total number of contexts.  

Moreover, users are connected by their friendship links, with which we can define the social graph of users:
\begin{definition}
\textbf{(Social Graph)}. Given users $\mathcal{U}$, the social graph is defined as $\mathcal{G}_{u}=\{\mathcal{U}, \mathcal{S}\}$, where $\mathcal{S}$ denotes edges.
For two users $u,v\in\mathcal{U}$, $ (u,v) \in \mathcal{S}$ if they are friends. 
\end{definition}

\begin{figure}[htpb]
         \centering
         \includegraphics[width=0.4\textwidth]{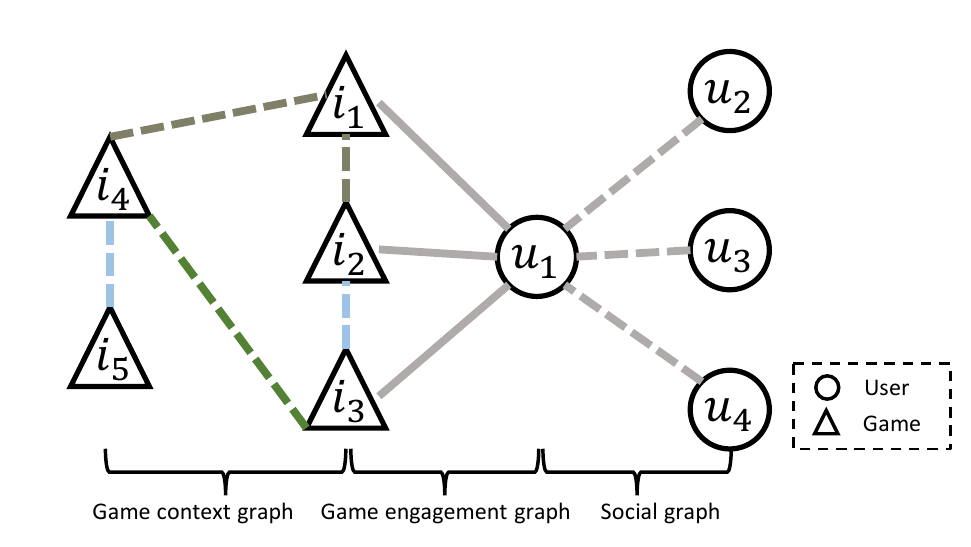}
         \caption{Graph structure of Steam data}
         \label{fig:graph}
\end{figure}

Figure~\ref{fig:graph} shows the graph structure. Game context graph and social graph act like side information to game engagement graph.

Thereafter, we can formulate game recommendations as follows:
\begin{definition}
\textbf{(Game Recommendation)}. Supposing a set of users $\mathcal{U}$ and games $\mathcal{I}$, given the game engagement of all users $\{\mathcal{E}_{u}\}|_{u\in\mathcal{U}}$, the game context graph $\mathcal{G}_i$ and the social graph $\mathcal{G}_{u}$, we should recommend a user $u\in\mathcal{U}$ a ranking list of games that $u$ has no engagements in.  
\end{definition}

\section{Proposed Model}
\begin{figure*}[htpb]
         \centering
         \includegraphics[width=0.9\textwidth]{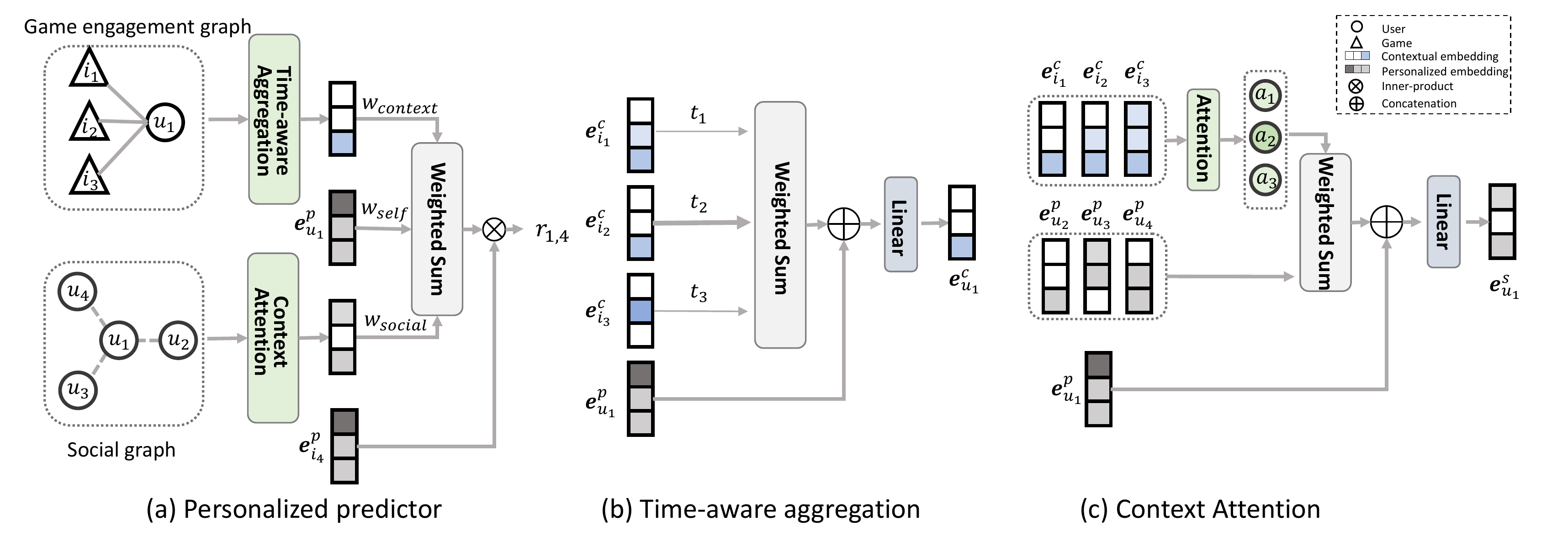}
         \caption{Framework of SCGRec}
         \label{fig:framework}
\end{figure*}

In this section, we introduce SCGRec. It has four crucial components: game context graph neural network, time-aware context aggregation, context-aware social aggregation, and the personalized predictor. The proposed model is shown in Figure~\ref{fig:framework}.

\subsection{Game context graph neural network}
Game context graph $\mathcal{G}_{i}$ is built to propagate information among similar games. In this dataset, we build $\mathcal{G}_{i}$ by $5$ game relations including $3$ feature-based relations and $2$ behavior-based relations. 

Feature-based relations include co-genre, co-developer, and co-publisher based on different game features. For example, edges in co-developer relation indicate the same game developer develops the connected games. Behavior-based relations are constructed from users' behavior information. Co-purchase relation measures whether two games are likely to be purchased together. The co-purchase score between game $i$ and $j$ is calculated by Equation~\ref{eq:co_purchase}. Then the edges with a co-purchase score larger than threshold $\tau_p$ remain in the co-purchase graph. The co-dwelling relation between two games reflects whether two games can attract users to play a similar period length. The co-dwelling score is computed by:
\begin{equation}
    \text{Co-dwelling} = e^{-\frac{|t_i - t_j|}{T}},
\end{equation}
where $t_i$ and $t_j$ are the average dwelling time for users co-purchased game $i$ and game $j$, respectively. $T$ is a constant to normalize the time. Then the edges with co-dwelling scores larger than threshold $\tau_t$ are kept in the co-dwelling graph.

\begin{figure}[htpb]
         \centering
         \includegraphics[width=0.35\textwidth]{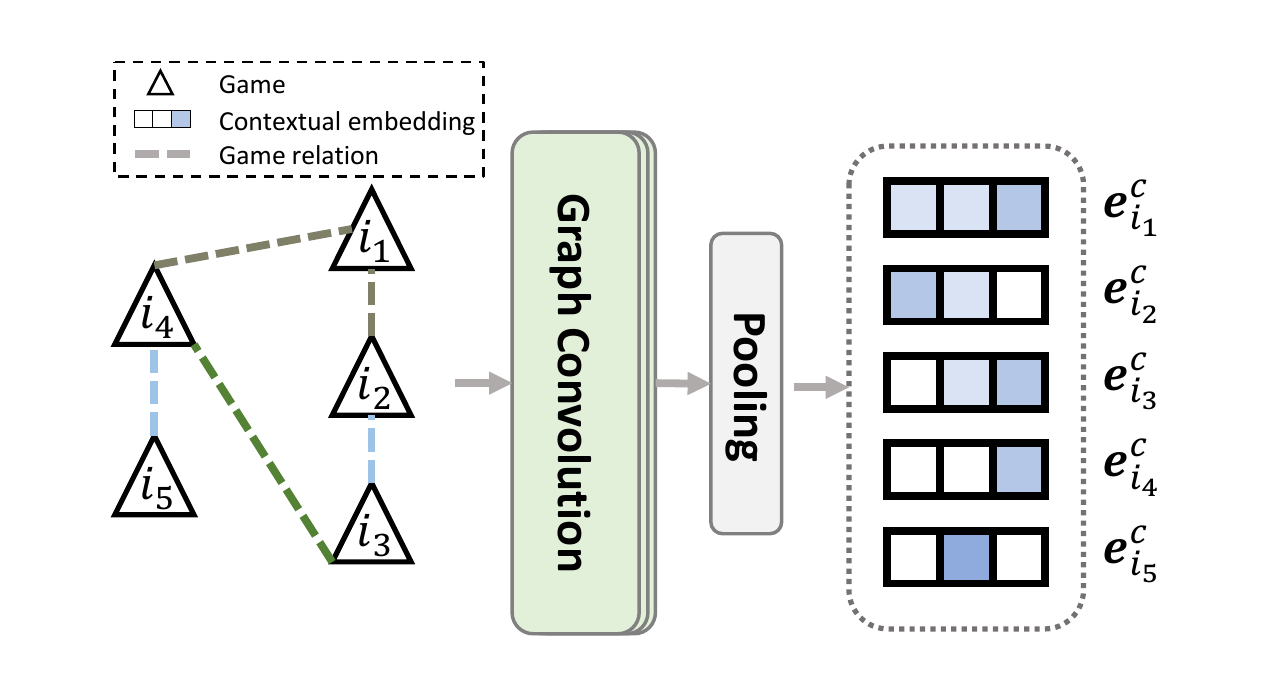}
         \caption{Game context graph aggregation}
         \label{fig:context_graph}
\end{figure}

As shown in Figure~\ref{fig:context_graph}, one graph convolution is performed for each relation type. Then the game contextual embedding $\mathbf{e}_{i}^{c}\in\mathbb{R}^{d}$ is obtained after a pooling layer to aggregate information from all relation types. Specifically, we use mean aggregation as:
\begin{equation}
    \mathbf{e}_{i}^{c} =\frac{1}{|C|}\sum_{c\in\mathcal{C}}\sum_{j\in\mathcal{N}^{c}(i)}\frac{1}{m_{i,j}^{c}}\mathbf{W}_{c}\mathbf{h}_j+\mathbf{b}_c,
\end{equation}
where $\mathcal{N}^{c}(i)$ is the neighbor set of node $i$ w.r.t. relation $c$. $\mathbf{h}_j$ denotes the hidden feature vector of node $j$. $m_{i,j}^{c}=\sqrt{|\mathcal{N}^c(i)|}\sqrt{|\mathcal{N}^c(i)|}$ is the normalization factor. $\mathbf{W}_{c}$ and $\mathbf{b}_{c}$ are graph convolution parameters.

\subsection{Time-aware context aggregation}
Dwelling time is a critical information for game recommendation, which signifies user engagements. 
However, time information can not be directly used because the dwelling time of different games are incomparable.
Instead, we calculate the user dwelling time percentile to measure user's engagement in each game. 

User's contextual embedding $\mathbf{e}_{u}^{c}\in\mathbb{R}^{d}$ is obtained from game contextual embedding $\mathbf{e}_{i}^{c}$ with the consideration of user engagement $\mathcal{E}_u$. For $u$, the time-aware context aggregation is calculated by:
\begin{equation}
    \mathbf{e}_{\text{agg}}^{c} = \sum_{(i,t)\in \mathcal{E}_u}\gamma_{i,t}\mathbf{e}_{i}^{c},
\end{equation}
where $\gamma_{i,t}$ is the time-aware weight, which is calculated as follows:
\begin{equation}
    \gamma_{i, t}=\frac{\text{percentile}(i,t)}{\sum_{(i',t')\in \mathcal{E}_u}\text{percentile}(i',t')},
\end{equation}
where $\text{percentile}(i,t)$ returns time $t$'s percentile in all dwelling time occurred in game $i$.

Then the contextual embedding of user $u$ is calculated by:
\begin{equation}
    \mathbf{e}_{u}^{c}=\mathbf{W}(\mathbf{e}_{u}^{p}\oplus\mathbf{e}_{\text{agg}}^{c}),
\end{equation}
where $\mathbf{W}\in\mathbb{R}^{d \times 2d}$ is a linear projection matrix, $\oplus$ is the concatenation operator and $\mathbf{e}_{u}^{p}$ is $u$'s personalized embedding.

\subsection{Context-aware social aggregation}
In Table~\ref{tab:Pearson_Correlation}, we observe higher correlation between friends than random users, which implies that social neighbors are beneficial for inferring users' interests. 
Since contextual embeddings of users contain the game engagement information, the aggregation of context embedding from social connection perspectives can help explore more similar neighbors. 
We use context attention to obtain the attention weight for social aggregation:
\begin{equation}
    \alpha_{u_i} = \mathrm{softmax_{u_j\in \mathcal{N}(u)}} (e_{u_i})=\frac{\exp \left(e_{u_i}\right)}{\sum_{u_j \in \mathcal{N}(u)} \exp \left(e_{u_j}\right)},
\end{equation}
\begin{equation}
    e_{u_i} = \mathrm{LeakyReLU}\left(\mathbf{a}^{\top} (\mathbf{W} \mathbf{e}_{u}^{c} \oplus \mathbf{W} \mathbf{e}_{u_i}^{c})\right),
\end{equation}
where $\mathbf{W}\in\mathbb{R}^{d \times d}$ and $\mathbf{a}\in\mathbb{R}^{2d}$ are parameters for calculating context attention weights. $\mathcal{N}(u)$ is the neighbor set of node $u$, $\mathbf{e}_{u}^{c}$ is node $u$'s contextual embedding, and $\oplus$ is the concatenation operator. We perform weighted aggregation on personalized embedding as:
\begin{equation}
        \mathbf{e}_{\text{agg}}^{p} = \sum_{u_i\in \mathcal{N}(u)}\alpha_{u_i}\mathbf{e}_{u_i}^{p},
\end{equation}
then the social embedding of user $u$ is calculated as:

\begin{equation}
    \mathbf{e}_{u}^{s}=\mathbf{W}(\mathbf{e}_{u}^{p}\oplus\mathbf{e}_{\text{agg}}^{p}),
\end{equation}
where $\mathbf{W}\in\mathbb{R}^{d \times 2d}$ is a linear transformation, $\oplus$ is the concatenation operator and $\mathbf{e}_{u}^{p}$ is $u$'s personalized embedding.

\subsection{Personalized Predictor}
For user $u$, we have context embedding $\mathbf{e}_{u}^{c}$ from time-aware context aggregation, social embedding $\mathbf{e}_{u}^{s}$ from context-aware social aggregation and personalized embedding $\mathbf{e}_{u}^{p}$. To simultaneously incorporating all these information, we assign different weights and sum them to obtain the final embedding of users as follows:
\begin{equation}
    \mathbf{e}_{u}=w_{\text{context}} \mathbf{e}_{u}^{c} + w_{\text{social}} \mathbf{e}_{u}^{s} + w_{\text{self}} \mathbf{e}_{u}^{p},
\end{equation}
where $w_{\text{context}}$, $w_{\text{social}}$ and $w_{\text{self}}$ are scalar hyper-parameters to decide the importance of the corresponding information. 
Hereafter, the final rating score is calculated by the dot-product between user embedding $\mathbf{e}_{u}$ and game personalized embedding $\mathbf{e}_{i}^{p}$ as follows:
\begin{equation}
    r_{u,i}=\mathbf{e}_{u}\cdot\mathbf{e}_{i}^{p}.
\end{equation}

We adopt the Bayesian Personalized Ranking (BPR) loss~\cite{DBLP:conf/uai/RendleFGS09} for optimizating all trainable parameters, which includes the embeddings and convolution weights. Loss function is defined as:
\begin{equation}
    \mathcal{L}=-\sum_{(u,i)\in \mathcal{E}_u, j\in \mathcal{I}\setminus\mathcal{E}_u}\log \sigma(r_{u,i}-r_{u,j}) + \lambda ||\Theta||^2_2,
\end{equation}
where $j$ denotes the negative games sampled from $\mathcal{I}\setminus\mathcal{E}_u$ that user $u$ has no interactions with, $\lambda$ is the hyper-parameter, and $\Theta$ includes all trainable parameters.

\section{Experiments}
\subsection{Experimental Setup}
\subsubsection{Steam Data Set}
Due to the limitation of computational resources, we preprocess the raw data by two steps to make it feasible for training and evaluation. 
Firstly, we filter out the users with less than $5$ game interactions or play games for less than $60$ minutes. Then $30\%$ of the users are randomly sampled to form the data set.
The details of the filtered dataset are shown in Table \ref{tab:data}. 
Given this dataset, We then build a validation set and a test set by randomly sampled $50,000$ players. 
The validation set and test set are formed by $10\%$ game interactions of these selected players. 
This processed data is publicly released\footnote{\url{https://drive.google.com/file/d/1F9kr_YWimBtexJEH-zkDzCOwl1q7GmFp/view?usp=sharing}} for future reference.

\begin{table}[htbp]
\caption{Statistics of filtered Steam Data}
\label{tab:data}
\begin{tabular}{lr}
\toprule
\midrule
\# Players &  3,908,744 \\
\# Games & 2,707 \\
\# Publishers & 689 \\
\# Developers & 1,170 \\
\# Interactions & 95,441,434 \\
\# Social Connections & 10,625,806 \\
\bottomrule 
\end{tabular} 
\end{table}

\subsubsection{Baselines}
To evaluate the intrinsic characteristics of game recommendation, we compare our proposed model with several baseline models, including Popularity (time), Popularity (count), LightGCN~\cite{he2020lightgcn}, RGCN~\cite{schlichtkrull2018modeling}, GIN~\cite{DBLP:conf/iclr/XuHLJ19}, PinSAGE~\cite{ying2018graph} and GAT~\cite{velivckovic2017graph}. Detailed description and comparison are given in the appendix.

\subsubsection{Implementation Details}
We implement SCGRec in Pytorch and deep graph library (DGL)\footnote{\url{https://www.dgl.ai/}}, an open-source framework for graph neural networks. Hyper-parameters include embedding size, learning rate, batch size, social embedding aggregation weight $W_{social}$ and contextural embedding aggregation weight $W_{context}$. The personalized embedding aggregation weight is set to $1-W_{social}-W_{context}$. The tuning range and best hyper-parameters setting is given in the appendix. Adam~\cite{DBLP:journals/corr/KingmaB14} is adopted as the optimizer. BPR loss is adopted for optimization with random negative sample. Early stop strategy is adopted to avoid over-fitting. The training is stopped if the model's performance does not increase in successive $10$ epochs.
During testing, we rank all the potential games from those 2,702 games for each user based on their interaction scores. Then we calculate the metrics including NDCG$@K$, Recall$@K$, Hit Ratio$@K$, and Precision$@K$, where $K$ ranges in $\{5,10,20\}$

\subsection{Performance Evaluation}

\begin{table*}[htbp]
\caption{Overall comparison, the best and second-best results are in bold and underlined, respectively}
\label{tab:comparison}
\begin{tabular}{lcccccccccccc}
\toprule

\multirow{2}{*}{Method} & \multicolumn{3}{c}{NDCG} & \multicolumn{3}{c}{Recall} & \multicolumn{3}{c}{Hit Ratio} & \multicolumn{3}{c}{Precision} \\

\cmidrule(r){2-4} \cmidrule(r){5-7} \cmidrule(r){8-10} \cmidrule{11-13}
& @5 & @10 & @20 & @5 & @10 & @20 & @5 & @10 & @20 & @5 & @10 & @20 \\
\midrule
Popularity (time) & 0.0896 & 0.1053 & 0.1256  & 0.1106 & 0.1584 & 0.2324 & 0.1600 & 0.2320 & 0.3290 & 0.0336 & 0.0255 & 0.0192 \\
Popularity (count) & 0.1654 & 0.2050 & 0.2380  & 0.2335 & 0.3527 & 0.4735 & 0.2960 & 0.4348 & 0.5745 & 0.0662 & 0.0512 & 0.0357 \\
LightGCN        & 0.1861 & 0.2100 & 0.2475 & 0.2452 & 0.3174 & 0.4507  & 0.2849 & 0.3784 & 0.5504  & 0.0637 & 0.0447 & 0.0347 \\
RGCN       & 0.2253 & 0.2616 & 0.2980 & 0.2888 & 0.4027 & 0.5359 & 0.3771 & 0.5107 & 0.6550 & 0.0863 & 0.0629 & 0.0441 \\
GIN        & 0.3824 & 0.4179 & 0.4460 & 0.4851 & 0.5924 & 0.6872 & 0.5978 & 0.7163 & 0.8060 & 0.1375 & 0.0912 & 0.0582 \\
PinSAGE  & 0.3963 & 0.4271 & 0.4529 & 0.4927 & 0.5863 & 0.6740 & 0.6025 & 0.7088 & 0.7926 & 0.1385 & 0.0899 & 0.0567 \\
GAT        & \underline{0.4053}  & \underline{0.4385} & \underline{0.4655} & \underline{0.5081} & \underline{0.6082} & \underline{0.6988} & \underline{0.6209} & \underline{0.7319} & \underline{0.8158} & \underline{0.1430}  & \underline{0.0936} & \underline{0.0591} \\
SCGRec & \textbf{0.4351} & \textbf{0.4660} & \textbf{0.4921} & \textbf{0.5385} & \textbf{0.6311} & \textbf{0.7175} & \textbf{0.6519} & \textbf{0.7535} & \textbf{0.8331} & \textbf{0.1508} & \textbf{0.0969} & \textbf{0.0609} \\
\hline
Improvement & 6.98\% & 6.25\% & 5.70\% & 5.96\% & 3.75\% & 2.66\% & 4.98\% & 2.94\% & 2.11\% & 5.43\% & 3.52\% & 2.93\% \\
\bottomrule 
\end{tabular} 
\end{table*}

 Experiment results are shown in Table~\ref{tab:comparison}. from the results. We have the following observations:
\begin{itemize}[leftmargin=*]
    \item SCGRec significantly outperforms the state-of-the-art GNN models. In the top $5$ recommendation, SCGRec outperforms the second-best model by more than $5\%$ on NDCG, Recall, and Precision. By tackling the personalization, social connection, and game interpretation information accordingly, SCGRec can effectively fuse all side information to enhance game recommendation. When the recommendation list gradually becomes longer, the improvement of SCGRec becomes more and more limited on all the metrics. For example, the improvement on Recall@5 is $5.96\%$ while on Recall@20 is only $2.66\%$. It shows the difficulty of improving on a long recommendation list.
    \item The personalized recommendation is much better than the non-personalized recommendation. Popularity (time) and Popularity (count) are non-personalized recommendation methods. They generate the same recommendation list for all users based on game popularity. The worse performance of popularity-based methods shows the importance of personalized recommendations. Popularity (count) performs nearly twice as well as popularity (time). It shows that directly utilizing dwelling time length is not informative. The same dwelling time reflects different engagements on different games. Thus, dwelling time information is modeled as user engagement percentile in SCGRec.
    \item Nodes' self-information is important in GNN aggregation. LightGCN and RGCN have much lower performance than other GNN based models. Compared to other GNN models, the main difference between LightGCN and RGCN is that they directly aggregate all the neighbor information without considering the relative importance of self-information. PinSage firstly aggregates neighbor embedding and self-embedding, then it fuses the information by a learnable linear layer. GIN also focuses more on self-embedding. It adds the aggregated neighbor embedding as side information. GAT learns attention weight for each neighbor with consideration on nodes' self-embedding. In SCGRec, we explicitly add nodes' self-information controlled by hyper-parameter $w_{\text{self}}$.
    \item Side information such as social links and game context graphs are helpful to improve recommendation performance. Compared with LightGCN, RGCN builds social link and game context information as different relation types and aggregates information from all relations. LightGCN only aggregates on a user-game bipartite graph. The improvement of RGCN over LightGCN can be explained by utilizing more side information. SCGRec explicitly models game-side information as contextual embedding by time-aware aggregation. Social link information is also modeled as social embedding by context-aware social aggregation module. 
    
\end{itemize}

\subsection{Ablation Study}
\begin{figure}
      \begin{center}
        \includegraphics[width=.22\textwidth]{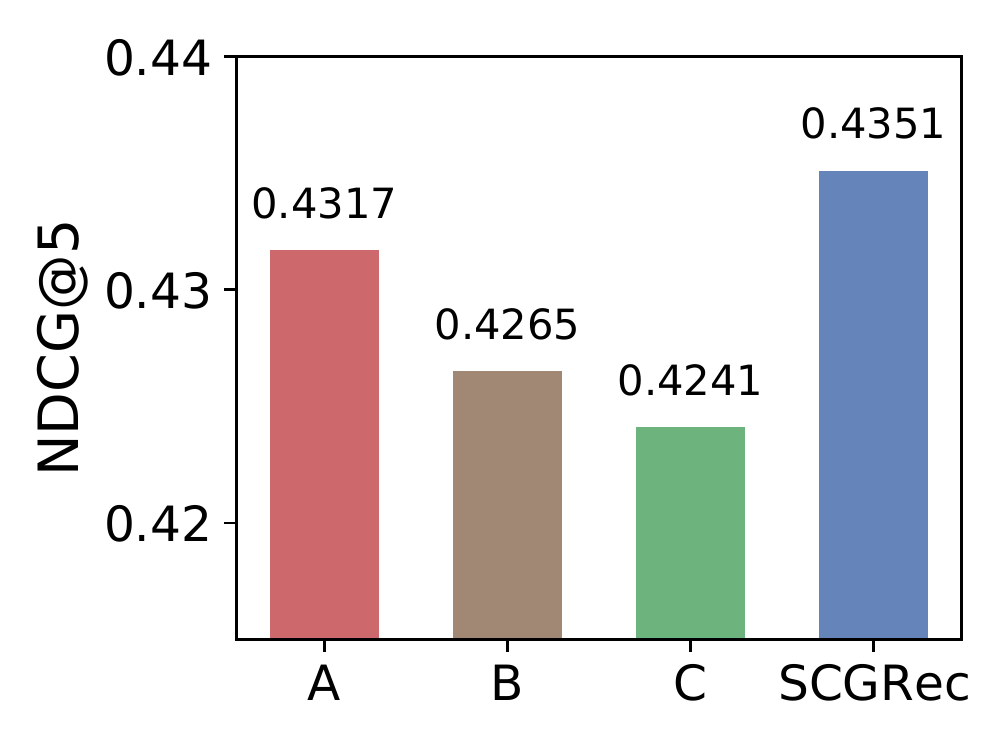}
        \includegraphics[width=.22\textwidth]{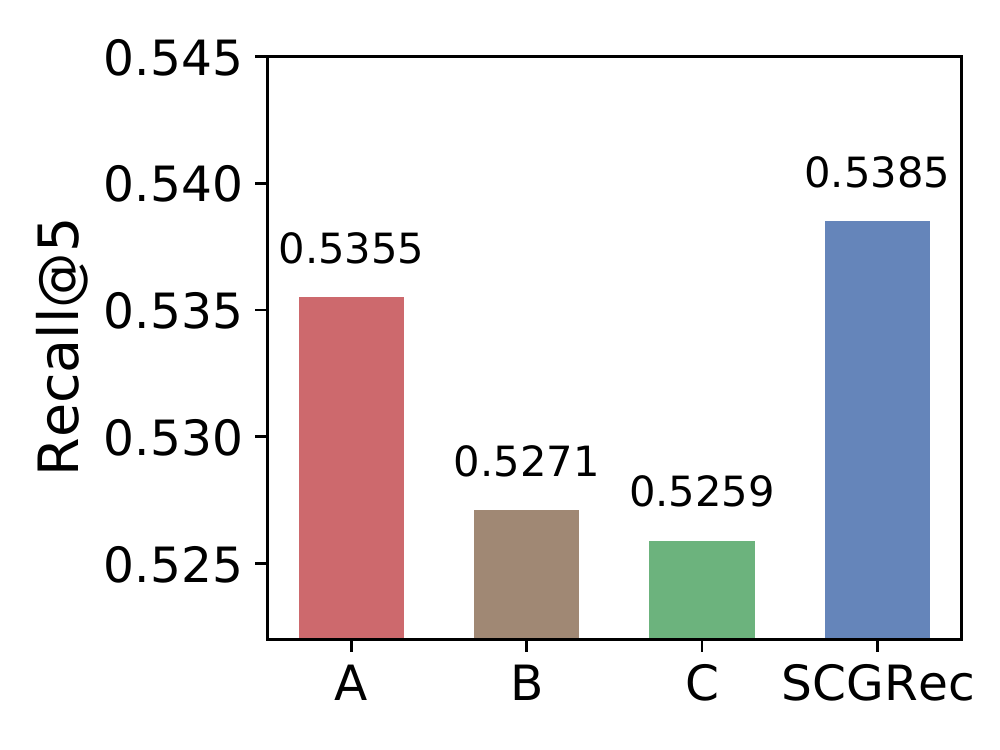}
      \end{center}
        \caption{Ablation study of SCGRec}
        \label{Ablation study}
\end{figure}

In this section, we perform an ablation study on SCGRec. We build three variant of SCGRec by removing part of its modules: 1) variant A is built by removing social embedding; 2) variant B removes those contextual embeddings; and 3) variant C is constructed by removing both social and contextual information. 
Experimental results are demonstrated in Figure~\ref{Ablation study}. 
Compared with all three variants, SCGRec performs the best on all metrics. 
We observe that considering both social embedding and contextual embedding achieves the best performance. 
Moreover, variant A performs better than variant B, which indicates that removing contextual embedding has more negative effects on SCGRec. 
Hence, the context information is relatively more important than social information. 
Compared with variant C, A and B both have better performance. 
This justifies the effectiveness of both the social connections and game contextualization in game recommendation. 

It is also worth noting that SCGRec performs better than the best baseline even without the side information. 
The reason is that the number of nodes in the game engagements is rather imbalanced, millions of players v.s. thousands of games. 
Hence, the direct neighbor aggregation of existing GNN methods on the player-game interaction graph leads to the over-smoothing problem. 
However, SCGRec adopts personalized embeddings, which are optimized directly from interaction rather than aggregation from neighbors. 

\subsection{Social and Context Influence}
\begin{figure}
      \begin{center}
        \includegraphics[width=.22\textwidth]{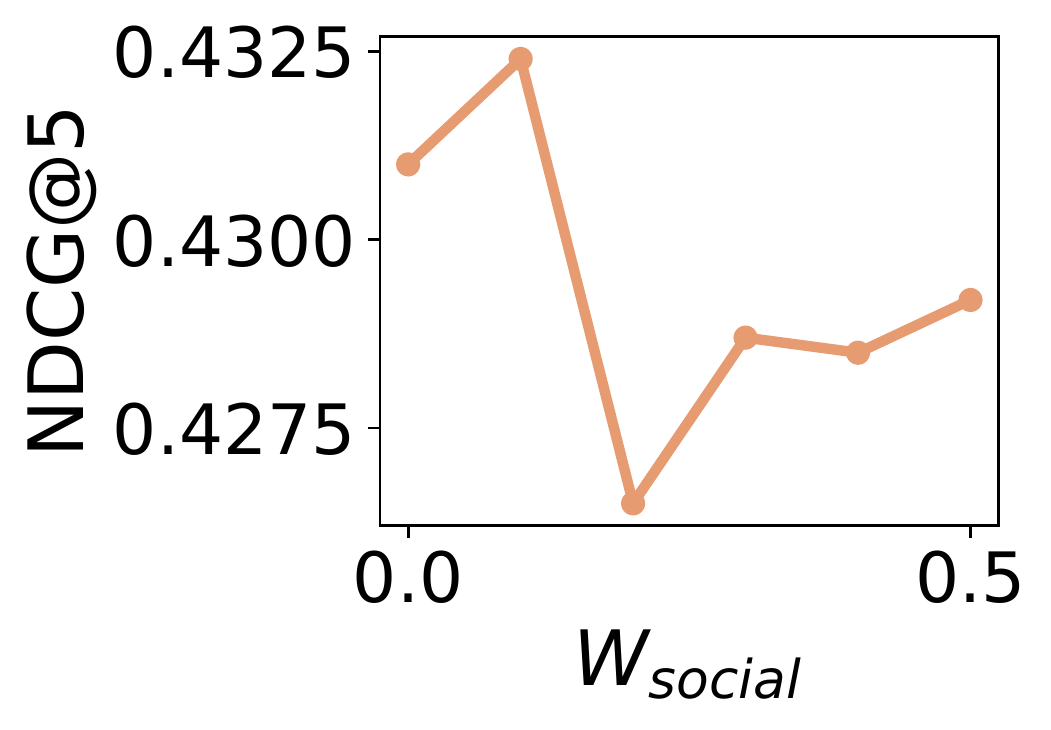}
        \includegraphics[width=.22\textwidth]{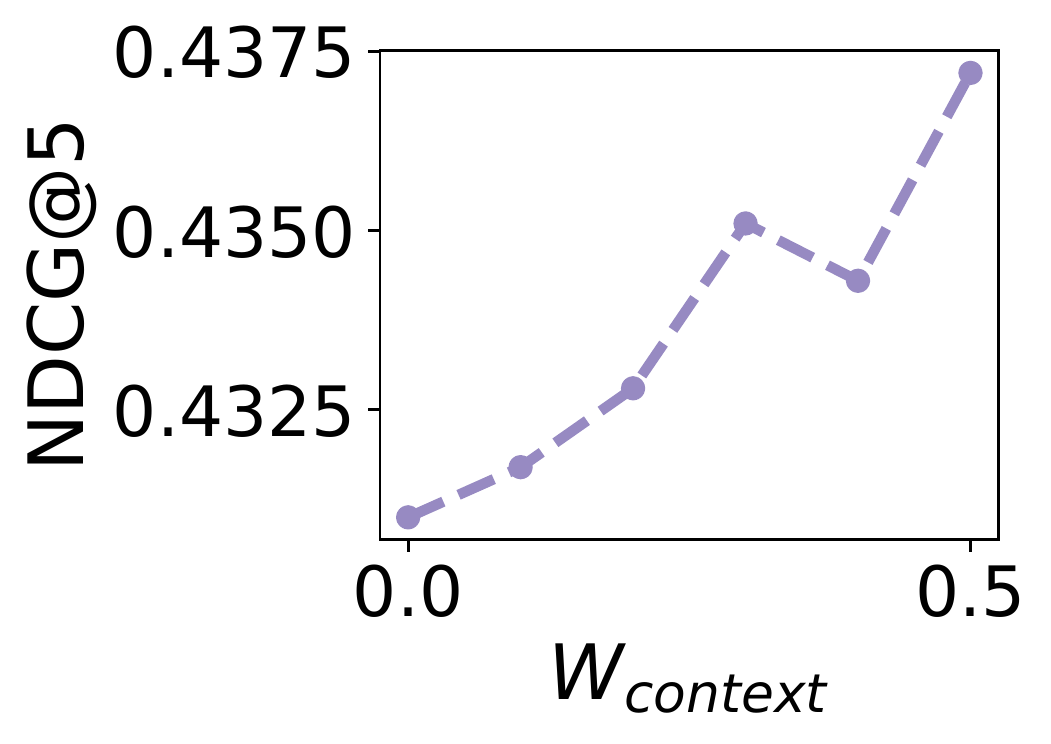}
      \end{center}
        \caption{Social and Context influence}
        \label{Influence analysis}
\end{figure}

In this section, we show the influence of social and context information by tuning the importance weight $w_{\text{social}}$ and $w_{\text{context}}$. $w_{\text{self}}$ is calculated by $w_{\text{self}}=1-w_{\text{social}}-w_{\text{context}}$.
Experiment results are shown in Figure~\ref{Influence analysis}. When $w_{\text{social}}$ gradually becomes larger, SCGRec's performance first increases and then decreases. It shows social information contains lots of noise. Similar observations are also shown in previous research. Junliang et al.~\cite{DBLP:conf/cikm/Yu0LYL18} discuss that explicit social links are not all reliable due to the existence of spammers and bots. Also, the inconsistent social neighbors~\cite{yang21consisrec} can also be seen as noise. When we pay much importance weight to social embedding, the noise in social neighbors is also amplified, which may lower down the performance. It can also be observed from Figure~\ref{Influence analysis} that when $w_{\text{context}}$ gradually becomes larger, SCGRec performs better. It shows the learned information from game context graph is beneficial. Providing the flexibility in constructing game context graph, SCGRec can easily model multi-type game context information to boost model's performance.

\section{Conclusion and Future Work}
In this paper, we comprehensively analyze nearly billion of user behaviors on the Steam online gaming platform, including dwelling time, social impacts, and game context information. 
Based on our analysis, we argue that $3$ key characteristics should be harnessed in the game recommendation, which are personalization, game contextualization and social connection.
Therefore, we propose a new model SCGRec by tackling those characteristics simultaneously for the game recommendation. 
Experimental results demonstrate the superiority of SCGRec regarding game recommendation performance. 
It has better performance against all existing GNN based methods in game recommendation. 
The SCGRec model is simple and rather flexible to incorporate various side information for recommendation, which
provides the game industry with an effective model to increase the game engagement of users.

We leave several research directions in game recommendations as to future work. Firstly, dwelling time measures user direct game engagement. But, how to effectively utilize dwelling time information is still an open question. 
Secondly, the user game engagement graph is a highly imbalanced bipartite graph. Differences in neighbor aggregation behavior on the two sides can be further investigated.
Lastly, as shown in data analyses, there exist a large portion of cold-start players. How to effectively recommend games for those players is yet another challenging research problem.   

\section{Acknowledgments}
This work was supported in part by NSF under grants III-1763325, III-1909323,  III-2106758, and SaTC-1930941.

\bibliographystyle{ACM-Reference-Format}
\bibliography{sample-base}


\begin{thebibliography}{56}


\ifx \showCODEN    \undefined \def \showCODEN     #1{\unskip}     \fi
\ifx \showDOI      \undefined \def \showDOI       #1{#1}\fi
\ifx \showISBNx    \undefined \def \showISBNx     #1{\unskip}     \fi
\ifx \showISBNxiii \undefined \def \showISBNxiii  #1{\unskip}     \fi
\ifx \showISSN     \undefined \def \showISSN      #1{\unskip}     \fi
\ifx \showLCCN     \undefined \def \showLCCN      #1{\unskip}     \fi
\ifx \shownote     \undefined \def \shownote      #1{#1}          \fi
\ifx \showarticletitle \undefined \def \showarticletitle #1{#1}   \fi
\ifx \showURL      \undefined \def \showURL       {\relax}        \fi
\providecommand\bibfield[2]{#2}
\providecommand\bibinfo[2]{#2}
\providecommand\natexlab[1]{#1}
\providecommand\showeprint[2][]{arXiv:#2}

\bibitem[\protect\citeauthoryear{Anwar, Shahzad, Sattar, Khan, and Majid}{Anwar
  et~al\mbox{.}}{2017}]%
        {anwar2017game}
\bibfield{author}{\bibinfo{person}{Syed~Muhammad Anwar}, \bibinfo{person}{Talha
  Shahzad}, \bibinfo{person}{Zunaira Sattar}, \bibinfo{person}{Rahma Khan},
  {and} \bibinfo{person}{Muhammad Majid}.} \bibinfo{year}{2017}\natexlab{}.
\newblock \showarticletitle{A game recommender system using collaborative
  filtering (GAMBIT)}. In \bibinfo{booktitle}{\emph{2017 14th International
  Bhurban Conference on Applied Sciences and Technology (IBCAST)}}. IEEE,
  \bibinfo{pages}{328--332}.
\newblock


\bibitem[\protect\citeauthoryear{Berg, Kipf, and Welling}{Berg
  et~al\mbox{.}}{2017}]%
        {berg2017graph}
\bibfield{author}{\bibinfo{person}{Rianne van~den Berg},
  \bibinfo{person}{Thomas~N Kipf}, {and} \bibinfo{person}{Max Welling}.}
  \bibinfo{year}{2017}\natexlab{}.
\newblock \showarticletitle{Graph convolutional matrix completion}.
\newblock \bibinfo{journal}{\emph{arXiv preprint arXiv:1706.02263}}
  (\bibinfo{year}{2017}).
\newblock


\bibitem[\protect\citeauthoryear{Bertens, Guitart, Chen, and
  Peri{\'a}{\~n}ez}{Bertens et~al\mbox{.}}{2018}]%
        {bertens2018machine}
\bibfield{author}{\bibinfo{person}{Paul Bertens}, \bibinfo{person}{Anna
  Guitart}, \bibinfo{person}{Pei~Pei Chen}, {and} \bibinfo{person}{{\'A}frica
  Peri{\'a}{\~n}ez}.} \bibinfo{year}{2018}\natexlab{}.
\newblock \showarticletitle{A machine-learning item recommendation system for
  video games}. In \bibinfo{booktitle}{\emph{2018 IEEE Conference on
  Computational Intelligence and Games (CIG)}}. IEEE, \bibinfo{pages}{1--4}.
\newblock


\bibitem[\protect\citeauthoryear{Chen and Gao}{Chen and Gao}{2018}]%
        {DBLP:journals/corr/abs-1803-08378}
\bibfield{author}{\bibinfo{person}{Lingjiao Chen} {and} \bibinfo{person}{Jian
  Gao}.} \bibinfo{year}{2018}\natexlab{}.
\newblock \showarticletitle{A trust-based recommendation method using network
  diffusion processes}.
\newblock \bibinfo{journal}{\emph{CoRR}}  \bibinfo{volume}{abs/1803.08378}
  (\bibinfo{year}{2018}).
\newblock
\showeprint[arXiv]{1803.08378}
\urldef\tempurl%
\url{http://arxiv.org/abs/1803.08378}
\showURL{%
\tempurl}


\bibitem[\protect\citeauthoryear{Cheuque, Guzm{\'a}n, and Parra}{Cheuque
  et~al\mbox{.}}{2019}]%
        {cheuque2019recommender}
\bibfield{author}{\bibinfo{person}{Germ{\'a}n Cheuque},
  \bibinfo{person}{Jos{\'e} Guzm{\'a}n}, {and} \bibinfo{person}{Denis Parra}.}
  \bibinfo{year}{2019}\natexlab{}.
\newblock \showarticletitle{Recommender systems for Online video game
  platforms: The case of STEAM}. In \bibinfo{booktitle}{\emph{Companion
  Proceedings of The 2019 World Wide Web Conference}}.
  \bibinfo{pages}{763--771}.
\newblock


\bibitem[\protect\citeauthoryear{De~Simone, Gadia, Maggiorini, and
  Ripamonti}{De~Simone et~al\mbox{.}}{2021}]%
        {de2021design}
\bibfield{author}{\bibinfo{person}{Lorenzo De~Simone}, \bibinfo{person}{Davide
  Gadia}, \bibinfo{person}{Dario Maggiorini}, {and} \bibinfo{person}{Laura~Anna
  Ripamonti}.} \bibinfo{year}{2021}\natexlab{}.
\newblock \showarticletitle{Design of a Recommender System for Video Games
  based on In-Game Player Profiling and Activities}. In
  \bibinfo{booktitle}{\emph{CHItaly 2021: 14th Biannual Conference of the
  Italian SIGCHI Chapter}}. \bibinfo{pages}{1--8}.
\newblock


\bibitem[\protect\citeauthoryear{Fan, Ma, Li, He, Zhao, Tang, and Yin}{Fan
  et~al\mbox{.}}{2019}]%
        {fan2019graph}
\bibfield{author}{\bibinfo{person}{Wenqi Fan}, \bibinfo{person}{Yao Ma},
  \bibinfo{person}{Qing Li}, \bibinfo{person}{Yuan He}, \bibinfo{person}{Eric
  Zhao}, \bibinfo{person}{Jiliang Tang}, {and} \bibinfo{person}{Dawei Yin}.}
  \bibinfo{year}{2019}\natexlab{}.
\newblock \showarticletitle{Graph neural networks for social recommendation}.
  In \bibinfo{booktitle}{\emph{The World Wide Web Conference, {WWW} 2019, San
  Francisco, CA, USA, May 13-17, 2019}}. \bibinfo{publisher}{{ACM}},
  \bibinfo{pages}{417--426}.
\newblock


\bibitem[\protect\citeauthoryear{Fan, Liu, Wang, Zheng, and Yu}{Fan
  et~al\mbox{.}}{2021a}]%
        {DBLP:conf/cikm/FanL00Y21}
\bibfield{author}{\bibinfo{person}{Ziwei Fan}, \bibinfo{person}{Zhiwei Liu},
  \bibinfo{person}{Shen Wang}, \bibinfo{person}{Lei Zheng}, {and}
  \bibinfo{person}{Philip~S. Yu}.} \bibinfo{year}{2021}\natexlab{a}.
\newblock \showarticletitle{Modeling Sequences as Distributions with
  Uncertainty for Sequential Recommendation}. In
  \bibinfo{booktitle}{\emph{{CIKM} '21: The 30th {ACM} International Conference
  on Information and Knowledge Management, Virtual Event, Queensland,
  Australia, November 1 - 5, 2021}},
  \bibfield{editor}{\bibinfo{person}{Gianluca Demartini},
  \bibinfo{person}{Guido Zuccon}, \bibinfo{person}{J.~Shane Culpepper},
  \bibinfo{person}{Zi~Huang}, {and} \bibinfo{person}{Hanghang Tong}} (Eds.).
  \bibinfo{publisher}{{ACM}}, \bibinfo{pages}{3019--3023}.
\newblock
\urldef\tempurl%
\url{https://doi.org/10.1145/3459637.3482145}
\showDOI{\tempurl}


\bibitem[\protect\citeauthoryear{Fan, Liu, Wang, Wang, Nazari, Zheng, Peng, and
  Yu}{Fan et~al\mbox{.}}{2022}]%
        {fan2022sequential}
\bibfield{author}{\bibinfo{person}{Ziwei Fan}, \bibinfo{person}{Zhiwei Liu},
  \bibinfo{person}{Yu Wang}, \bibinfo{person}{Alice Wang},
  \bibinfo{person}{Zahra Nazari}, \bibinfo{person}{Lei Zheng},
  \bibinfo{person}{Hao Peng}, {and} \bibinfo{person}{Philip~S Yu}.}
  \bibinfo{year}{2022}\natexlab{}.
\newblock \showarticletitle{Sequential Recommendation via Stochastic
  Self-Attention}.
\newblock \bibinfo{journal}{\emph{arXiv preprint arXiv:2201.06035}}
  (\bibinfo{year}{2022}).
\newblock


\bibitem[\protect\citeauthoryear{Fan, Liu, Zhang, Xiong, Zheng, and Yu}{Fan
  et~al\mbox{.}}{2021b}]%
        {DBLP:conf/cikm/FanLZX0Y21}
\bibfield{author}{\bibinfo{person}{Ziwei Fan}, \bibinfo{person}{Zhiwei Liu},
  \bibinfo{person}{Jiawei Zhang}, \bibinfo{person}{Yun Xiong},
  \bibinfo{person}{Lei Zheng}, {and} \bibinfo{person}{Philip~S. Yu}.}
  \bibinfo{year}{2021}\natexlab{b}.
\newblock \showarticletitle{Continuous-Time Sequential Recommendation with
  Temporal Graph Collaborative Transformer}. In
  \bibinfo{booktitle}{\emph{{CIKM} '21: The 30th {ACM} International Conference
  on Information and Knowledge Management, Virtual Event, Queensland,
  Australia, November 1 - 5, 2021}},
  \bibfield{editor}{\bibinfo{person}{Gianluca Demartini},
  \bibinfo{person}{Guido Zuccon}, \bibinfo{person}{J.~Shane Culpepper},
  \bibinfo{person}{Zi~Huang}, {and} \bibinfo{person}{Hanghang Tong}} (Eds.).
  \bibinfo{publisher}{{ACM}}, \bibinfo{pages}{433--442}.
\newblock
\urldef\tempurl%
\url{https://doi.org/10.1145/3459637.3482242}
\showDOI{\tempurl}


\bibitem[\protect\citeauthoryear{Guo, Yang, Chen, Chen, Gao, and Ma}{Guo
  et~al\mbox{.}}{2020}]%
        {DBLP:journals/entropy/GuoYCCGM20}
\bibfield{author}{\bibinfo{person}{Chungu Guo}, \bibinfo{person}{Liangwei
  Yang}, \bibinfo{person}{Xiao Chen}, \bibinfo{person}{Duanbing Chen},
  \bibinfo{person}{Hui Gao}, {and} \bibinfo{person}{Jing Ma}.}
  \bibinfo{year}{2020}\natexlab{}.
\newblock \showarticletitle{Influential Nodes Identification in Complex
  Networks via Information Entropy}.
\newblock \bibinfo{journal}{\emph{Entropy}} \bibinfo{volume}{22},
  \bibinfo{number}{2} (\bibinfo{year}{2020}), \bibinfo{pages}{242}.
\newblock
\urldef\tempurl%
\url{https://doi.org/10.3390/e22020242}
\showDOI{\tempurl}


\bibitem[\protect\citeauthoryear{Guo, Tang, Ye, Li, and He}{Guo
  et~al\mbox{.}}{2017}]%
        {DBLP:conf/ijcai/GuoTYLH17}
\bibfield{author}{\bibinfo{person}{Huifeng Guo}, \bibinfo{person}{Ruiming
  Tang}, \bibinfo{person}{Yunming Ye}, \bibinfo{person}{Zhenguo Li}, {and}
  \bibinfo{person}{Xiuqiang He}.} \bibinfo{year}{2017}\natexlab{}.
\newblock \showarticletitle{DeepFM: {A} Factorization-Machine based Neural
  Network for {CTR} Prediction}. In \bibinfo{booktitle}{\emph{Proceedings of
  the Twenty-Sixth International Joint Conference on Artificial Intelligence,
  {IJCAI} 2017, Melbourne, Australia, August 19-25, 2017}},
  \bibfield{editor}{\bibinfo{person}{Carles Sierra}} (Ed.).
  \bibinfo{publisher}{ijcai.org}, \bibinfo{pages}{1725--1731}.
\newblock
\urldef\tempurl%
\url{https://doi.org/10.24963/ijcai.2017/239}
\showDOI{\tempurl}


\bibitem[\protect\citeauthoryear{He, Balasubramanian, Ceyani, Yang, Xie, Sun,
  He, Yang, Yu, Rong, et~al\mbox{.}}{He et~al\mbox{.}}{2021}]%
        {he2021fedgraphnn}
\bibfield{author}{\bibinfo{person}{Chaoyang He}, \bibinfo{person}{Keshav
  Balasubramanian}, \bibinfo{person}{Emir Ceyani}, \bibinfo{person}{Carl Yang},
  \bibinfo{person}{Han Xie}, \bibinfo{person}{Lichao Sun},
  \bibinfo{person}{Lifang He}, \bibinfo{person}{Liangwei Yang},
  \bibinfo{person}{Philip~S Yu}, \bibinfo{person}{Yu Rong}, {et~al\mbox{.}}}
  \bibinfo{year}{2021}\natexlab{}.
\newblock \showarticletitle{Fedgraphnn: A federated learning system and
  benchmark for graph neural networks}.
\newblock \bibinfo{journal}{\emph{arXiv preprint arXiv:2104.07145}}
  (\bibinfo{year}{2021}).
\newblock


\bibitem[\protect\citeauthoryear{He, Deng, Wang, Li, Zhang, and Wang}{He
  et~al\mbox{.}}{2020}]%
        {he2020lightgcn}
\bibfield{author}{\bibinfo{person}{Xiangnan He}, \bibinfo{person}{Kuan Deng},
  \bibinfo{person}{Xiang Wang}, \bibinfo{person}{Yan Li},
  \bibinfo{person}{Yongdong Zhang}, {and} \bibinfo{person}{Meng Wang}.}
  \bibinfo{year}{2020}\natexlab{}.
\newblock \showarticletitle{Lightgcn: Simplifying and powering graph
  convolution network for recommendation}. In
  \bibinfo{booktitle}{\emph{Proceedings of the 43rd International ACM SIGIR
  conference on research and development in Information Retrieval}}.
  \bibinfo{pages}{639--648}.
\newblock


\bibitem[\protect\citeauthoryear{Kim, Wi, Jang, and Kim}{Kim
  et~al\mbox{.}}{2020}]%
        {kim2020sequential}
\bibfield{author}{\bibinfo{person}{JaeWon Kim}, \bibinfo{person}{JeongA Wi},
  \bibinfo{person}{SooJin Jang}, {and} \bibinfo{person}{YoungBin Kim}.}
  \bibinfo{year}{2020}\natexlab{}.
\newblock \showarticletitle{Sequential Recommendations on Board-Game
  Platforms}.
\newblock \bibinfo{journal}{\emph{Symmetry}} \bibinfo{volume}{12},
  \bibinfo{number}{2} (\bibinfo{year}{2020}), \bibinfo{pages}{210}.
\newblock


\bibitem[\protect\citeauthoryear{Kingma and Ba}{Kingma and Ba}{2015}]%
        {DBLP:journals/corr/KingmaB14}
\bibfield{author}{\bibinfo{person}{Diederik~P. Kingma} {and}
  \bibinfo{person}{Jimmy Ba}.} \bibinfo{year}{2015}\natexlab{}.
\newblock \showarticletitle{Adam: {A} Method for Stochastic Optimization}. In
  \bibinfo{booktitle}{\emph{3rd International Conference on Learning
  Representations, {ICLR} 2015, San Diego, CA, USA, May 7-9, 2015, Conference
  Track Proceedings}}, \bibfield{editor}{\bibinfo{person}{Yoshua Bengio} {and}
  \bibinfo{person}{Yann LeCun}} (Eds.).
\newblock
\urldef\tempurl%
\url{http://arxiv.org/abs/1412.6980}
\showURL{%
\tempurl}


\bibitem[\protect\citeauthoryear{Kipf and Welling}{Kipf and Welling}{2017}]%
        {DBLP:conf/iclr/KipfW17}
\bibfield{author}{\bibinfo{person}{Thomas~N. Kipf} {and} \bibinfo{person}{Max
  Welling}.} \bibinfo{year}{2017}\natexlab{}.
\newblock \showarticletitle{Semi-Supervised Classification with Graph
  Convolutional Networks}. In \bibinfo{booktitle}{\emph{5th International
  Conference on Learning Representations, {ICLR} 2017, Toulon, France, April
  24-26, 2017, Conference Track Proceedings}}.
  \bibinfo{publisher}{OpenReview.net}.
\newblock
\urldef\tempurl%
\url{https://openreview.net/forum?id=SJU4ayYgl}
\showURL{%
\tempurl}


\bibitem[\protect\citeauthoryear{Liu, Fan, Wang, and Yu}{Liu
  et~al\mbox{.}}{2021a}]%
        {10.1145/3404835.3463036}
\bibfield{author}{\bibinfo{person}{Zhiwei Liu}, \bibinfo{person}{Ziwei Fan},
  \bibinfo{person}{Yu Wang}, {and} \bibinfo{person}{Philip~S. Yu}.}
  \bibinfo{year}{2021}\natexlab{a}.
\newblock \bibinfo{booktitle}{\emph{Augmenting Sequential Recommendation with
  Pseudo-Prior Items via Reversely Pre-Training Transformer}}.
\newblock \bibinfo{publisher}{Association for Computing Machinery},
  \bibinfo{address}{New York, NY, USA}, \bibinfo{pages}{1608–1612}.
\newblock
\showISBNx{9781450380379}
\urldef\tempurl%
\url{https://doi.org/10.1145/3404835.3463036}
\showURL{%
\tempurl}


\bibitem[\protect\citeauthoryear{Liu, Li, Fan, Guo, Achan, and Philip}{Liu
  et~al\mbox{.}}{2020a}]%
        {liu2020basket}
\bibfield{author}{\bibinfo{person}{Zhiwei Liu}, \bibinfo{person}{Xiaohan Li},
  \bibinfo{person}{Ziwei Fan}, \bibinfo{person}{Stephen Guo},
  \bibinfo{person}{Kannan Achan}, {and} \bibinfo{person}{S~Yu Philip}.}
  \bibinfo{year}{2020}\natexlab{a}.
\newblock \showarticletitle{Basket recommendation with multi-intent translation
  graph neural network}. In \bibinfo{booktitle}{\emph{2020 IEEE International
  Conference on Big Data (Big Data)}}. IEEE, \bibinfo{pages}{728--737}.
\newblock


\bibitem[\protect\citeauthoryear{Liu, Wan, Guo, Achan, and Yu}{Liu
  et~al\mbox{.}}{2020b}]%
        {liu2020basconv}
\bibfield{author}{\bibinfo{person}{Zhiwei Liu}, \bibinfo{person}{Mengting Wan},
  \bibinfo{person}{Stephen Guo}, \bibinfo{person}{Kannan Achan}, {and}
  \bibinfo{person}{Philip~S Yu}.} \bibinfo{year}{2020}\natexlab{b}.
\newblock \showarticletitle{Basconv: Aggregating heterogeneous interactions for
  basket recommendation with graph convolutional neural network}. In
  \bibinfo{booktitle}{\emph{Proceedings of the 2020 SIAM International
  Conference on Data Mining}}. SIAM, \bibinfo{pages}{64--72}.
\newblock


\bibitem[\protect\citeauthoryear{Liu, Yang, Fan, Peng, and Yu}{Liu
  et~al\mbox{.}}{2021b}]%
        {liu2021federated}
\bibfield{author}{\bibinfo{person}{Zhiwei Liu}, \bibinfo{person}{Liangwei
  Yang}, \bibinfo{person}{Ziwei Fan}, \bibinfo{person}{Hao Peng}, {and}
  \bibinfo{person}{Philip~S Yu}.} \bibinfo{year}{2021}\natexlab{b}.
\newblock \showarticletitle{Federated Social Recommendation with Graph Neural
  Network}.
\newblock \bibinfo{journal}{\emph{arXiv preprint arXiv:2111.10778}}
  (\bibinfo{year}{2021}).
\newblock


\bibitem[\protect\citeauthoryear{Machado, Gopstein, Nealen, and
  Togelius}{Machado et~al\mbox{.}}{2019}]%
        {machado2019pitako}
\bibfield{author}{\bibinfo{person}{Tiago Machado}, \bibinfo{person}{Dan
  Gopstein}, \bibinfo{person}{Andy Nealen}, {and} \bibinfo{person}{Julian
  Togelius}.} \bibinfo{year}{2019}\natexlab{}.
\newblock \showarticletitle{Pitako-recommending game design elements in
  cicero}. In \bibinfo{booktitle}{\emph{2019 IEEE Conference on Games (CoG)}}.
  IEEE, \bibinfo{pages}{1--8}.
\newblock


\bibitem[\protect\citeauthoryear{Mao, Zhu, Xiao, Lu, Wang, and He}{Mao
  et~al\mbox{.}}{2021}]%
        {mao2021ultragcn}
\bibfield{author}{\bibinfo{person}{Kelong Mao}, \bibinfo{person}{Jieming Zhu},
  \bibinfo{person}{Xi Xiao}, \bibinfo{person}{Biao Lu},
  \bibinfo{person}{Zhaowei Wang}, {and} \bibinfo{person}{Xiuqiang He}.}
  \bibinfo{year}{2021}\natexlab{}.
\newblock \showarticletitle{UltraGCN: Ultra Simplification of Graph
  Convolutional Networks for Recommendation}. In
  \bibinfo{booktitle}{\emph{Proceedings of the 30th ACM International
  Conference on Information \& Knowledge Management}}.
  \bibinfo{pages}{1253--1262}.
\newblock


\bibitem[\protect\citeauthoryear{Marchand and Hennig-Thurau}{Marchand and
  Hennig-Thurau}{2013}]%
        {marchand2013value}
\bibfield{author}{\bibinfo{person}{Andr{\'e} Marchand} {and}
  \bibinfo{person}{Thorsten Hennig-Thurau}.} \bibinfo{year}{2013}\natexlab{}.
\newblock \showarticletitle{Value creation in the video game industry: Industry
  economics, consumer benefits, and research opportunities}.
\newblock \bibinfo{journal}{\emph{Journal of interactive marketing}}
  \bibinfo{volume}{27}, \bibinfo{number}{3} (\bibinfo{year}{2013}),
  \bibinfo{pages}{141--157}.
\newblock


\bibitem[\protect\citeauthoryear{Mensah, Hui, and Yang}{Mensah
  et~al\mbox{.}}{2020}]%
        {DBLP:journals/algorithms/MensahGY20}
\bibfield{author}{\bibinfo{person}{Dennis Nii~Ayeh Mensah},
  \bibinfo{person}{Gao Hui}, {and} \bibinfo{person}{Liangwei Yang}.}
  \bibinfo{year}{2020}\natexlab{}.
\newblock \showarticletitle{Approximation Algorithm for Shortest Path in Large
  Social Networks}.
\newblock \bibinfo{journal}{\emph{Algorithms}} \bibinfo{volume}{13},
  \bibinfo{number}{2} (\bibinfo{year}{2020}), \bibinfo{pages}{36}.
\newblock
\urldef\tempurl%
\url{https://doi.org/10.3390/a13020036}
\showDOI{\tempurl}


\bibitem[\protect\citeauthoryear{O'Neill, Vaziripour, Wu, and Zappala}{O'Neill
  et~al\mbox{.}}{[n.\,d.]}]%
        {DBLP:conf/imc/ONeillVWZ16}
\bibfield{author}{\bibinfo{person}{Mark O'Neill}, \bibinfo{person}{Elham
  Vaziripour}, \bibinfo{person}{Justin Wu}, {and} \bibinfo{person}{Daniel
  Zappala}.} \bibinfo{year}{[n.\,d.]}\natexlab{}.
\newblock \showarticletitle{Condensing Steam: Distilling the Diversity of Gamer
  Behavior}. In \bibinfo{booktitle}{\emph{Proceedings of the 2016 {ACM} on
  Internet Measurement Conference, {IMC} 2016, Santa Monica, CA, USA, November
  14-16, 2016}}, \bibfield{editor}{\bibinfo{person}{Phillipa Gill},
  \bibinfo{person}{John~S. Heidemann}, \bibinfo{person}{John~W. Byers}, {and}
  \bibinfo{person}{Ramesh Govindan}} (Eds.). \bibinfo{pages}{81--95}.
\newblock


\bibitem[\protect\citeauthoryear{Pazzani and Billsus}{Pazzani and
  Billsus}{2007}]%
        {pazzani2007content}
\bibfield{author}{\bibinfo{person}{Michael~J Pazzani} {and}
  \bibinfo{person}{Daniel Billsus}.} \bibinfo{year}{2007}\natexlab{}.
\newblock \showarticletitle{Content-based recommendation systems}.
\newblock In \bibinfo{booktitle}{\emph{The adaptive web}}.
  \bibinfo{publisher}{Springer}, \bibinfo{pages}{325--341}.
\newblock


\bibitem[\protect\citeauthoryear{P{\'e}rez-Marcos, Mart{\'\i}n-G{\'o}mez,
  Jim{\'e}nez-Bravo, L{\'o}pez, and Moreno-Garc{\'\i}a}{P{\'e}rez-Marcos
  et~al\mbox{.}}{2020}]%
        {perez2020hybrid}
\bibfield{author}{\bibinfo{person}{Javier P{\'e}rez-Marcos},
  \bibinfo{person}{Luc{\'\i}a Mart{\'\i}n-G{\'o}mez}, \bibinfo{person}{Diego~M
  Jim{\'e}nez-Bravo}, \bibinfo{person}{Vivian~F L{\'o}pez}, {and}
  \bibinfo{person}{Mar{\'\i}a~N Moreno-Garc{\'\i}a}.}
  \bibinfo{year}{2020}\natexlab{}.
\newblock \showarticletitle{Hybrid system for video game recommendation based
  on implicit ratings and social networks}.
\newblock \bibinfo{journal}{\emph{Journal of Ambient Intelligence and Humanized
  Computing}} \bibinfo{volume}{11}, \bibinfo{number}{11}
  (\bibinfo{year}{2020}), \bibinfo{pages}{4525--4535}.
\newblock


\bibitem[\protect\citeauthoryear{Rendle}{Rendle}{2012}]%
        {rendle2012factorization}
\bibfield{author}{\bibinfo{person}{Steffen Rendle}.}
  \bibinfo{year}{2012}\natexlab{}.
\newblock \showarticletitle{Factorization machines with libfm}.
\newblock \bibinfo{journal}{\emph{ACM Transactions on Intelligent Systems and
  Technology (TIST)}} \bibinfo{volume}{3}, \bibinfo{number}{3}
  (\bibinfo{year}{2012}), \bibinfo{pages}{1--22}.
\newblock


\bibitem[\protect\citeauthoryear{Rendle, Freudenthaler, Gantner, and
  Schmidt{-}Thieme}{Rendle et~al\mbox{.}}{2009}]%
        {DBLP:conf/uai/RendleFGS09}
\bibfield{author}{\bibinfo{person}{Steffen Rendle}, \bibinfo{person}{Christoph
  Freudenthaler}, \bibinfo{person}{Zeno Gantner}, {and} \bibinfo{person}{Lars
  Schmidt{-}Thieme}.} \bibinfo{year}{2009}\natexlab{}.
\newblock \showarticletitle{{BPR:} Bayesian Personalized Ranking from Implicit
  Feedback}. In \bibinfo{booktitle}{\emph{{UAI} 2009, Proceedings of the
  Twenty-Fifth Conference on Uncertainty in Artificial Intelligence, Montreal,
  QC, Canada, June 18-21, 2009}}, \bibfield{editor}{\bibinfo{person}{Jeff~A.
  Bilmes} {and} \bibinfo{person}{Andrew~Y. Ng}} (Eds.).
  \bibinfo{publisher}{{AUAI} Press}, \bibinfo{pages}{452--461}.
\newblock
\urldef\tempurl%
\url{https://dslpitt.org/uai/displayArticleDetails.jsp?mmnu=1\&smnu=2\&article\_id=1630\&proceeding\_id=25}
\showURL{%
\tempurl}


\bibitem[\protect\citeauthoryear{Schlichtkrull, Kipf, Bloem, Van Den~Berg,
  Titov, and Welling}{Schlichtkrull et~al\mbox{.}}{2018}]%
        {schlichtkrull2018modeling}
\bibfield{author}{\bibinfo{person}{Michael Schlichtkrull},
  \bibinfo{person}{Thomas~N Kipf}, \bibinfo{person}{Peter Bloem},
  \bibinfo{person}{Rianne Van Den~Berg}, \bibinfo{person}{Ivan Titov}, {and}
  \bibinfo{person}{Max Welling}.} \bibinfo{year}{2018}\natexlab{}.
\newblock \showarticletitle{Modeling relational data with graph convolutional
  networks}. In \bibinfo{booktitle}{\emph{European semantic web conference}}.
  Springer, \bibinfo{pages}{593--607}.
\newblock


\bibitem[\protect\citeauthoryear{Shen, Wu, Zhang, Shan, Zhang, Letaief, and
  Li}{Shen et~al\mbox{.}}{2021}]%
        {shen2021powerful}
\bibfield{author}{\bibinfo{person}{Yifei Shen}, \bibinfo{person}{Yongji Wu},
  \bibinfo{person}{Yao Zhang}, \bibinfo{person}{Caihua Shan},
  \bibinfo{person}{Jun Zhang}, \bibinfo{person}{B~Khaled Letaief}, {and}
  \bibinfo{person}{Dongsheng Li}.} \bibinfo{year}{2021}\natexlab{}.
\newblock \showarticletitle{How Powerful is Graph Convolution for
  Recommendation?}. In \bibinfo{booktitle}{\emph{Proceedings of the 30th ACM
  International Conference on Information \& Knowledge Management}}.
  \bibinfo{pages}{1619--1629}.
\newblock


\bibitem[\protect\citeauthoryear{Sifa, Bauckhage, and Drachen}{Sifa
  et~al\mbox{.}}{2014a}]%
        {sifa2014archetypal}
\bibfield{author}{\bibinfo{person}{Rafet Sifa}, \bibinfo{person}{Christian
  Bauckhage}, {and} \bibinfo{person}{Anders Drachen}.}
  \bibinfo{year}{2014}\natexlab{a}.
\newblock \showarticletitle{Archetypal Game Recommender Systems.}. In
  \bibinfo{booktitle}{\emph{LWA}}. \bibinfo{pages}{45--56}.
\newblock


\bibitem[\protect\citeauthoryear{Sifa, Bauckhage, and Drachen}{Sifa
  et~al\mbox{.}}{2014b}]%
        {sifa2014playtime}
\bibfield{author}{\bibinfo{person}{Rafet Sifa}, \bibinfo{person}{Christian
  Bauckhage}, {and} \bibinfo{person}{Anders Drachen}.}
  \bibinfo{year}{2014}\natexlab{b}.
\newblock \showarticletitle{The Playtime Principle: Large-scale cross-games
  interest modeling}. In \bibinfo{booktitle}{\emph{2014 IEEE conference on
  computational intelligence and games}}. IEEE, \bibinfo{pages}{1--8}.
\newblock


\bibitem[\protect\citeauthoryear{Su and Khoshgoftaar}{Su and
  Khoshgoftaar}{2009}]%
        {DBLP:journals/advai/SuK09}
\bibfield{author}{\bibinfo{person}{Xiaoyuan Su} {and} \bibinfo{person}{Taghi~M.
  Khoshgoftaar}.} \bibinfo{year}{2009}\natexlab{}.
\newblock \showarticletitle{A Survey of Collaborative Filtering Techniques}.
\newblock \bibinfo{journal}{\emph{Adv. Artif. Intell.}}  \bibinfo{volume}{2009}
  (\bibinfo{year}{2009}), \bibinfo{pages}{421425:1--421425:19}.
\newblock
\urldef\tempurl%
\url{https://doi.org/10.1155/2009/421425}
\showDOI{\tempurl}


\bibitem[\protect\citeauthoryear{Tak{\'a}cs and Tikk}{Tak{\'a}cs and
  Tikk}{2012}]%
        {takacs2012alternating}
\bibfield{author}{\bibinfo{person}{G{\'a}bor Tak{\'a}cs} {and}
  \bibinfo{person}{Domonkos Tikk}.} \bibinfo{year}{2012}\natexlab{}.
\newblock \showarticletitle{Alternating least squares for personalized
  ranking}. In \bibinfo{booktitle}{\emph{Proceedings of the sixth ACM
  conference on Recommender systems}}. \bibinfo{pages}{83--90}.
\newblock


\bibitem[\protect\citeauthoryear{Veli{\v{c}}kovi{\'c}, Cucurull, Casanova,
  Romero, Lio, and Bengio}{Veli{\v{c}}kovi{\'c} et~al\mbox{.}}{2017}]%
        {velivckovic2017graph}
\bibfield{author}{\bibinfo{person}{Petar Veli{\v{c}}kovi{\'c}},
  \bibinfo{person}{Guillem Cucurull}, \bibinfo{person}{Arantxa Casanova},
  \bibinfo{person}{Adriana Romero}, \bibinfo{person}{Pietro Lio}, {and}
  \bibinfo{person}{Yoshua Bengio}.} \bibinfo{year}{2017}\natexlab{}.
\newblock \showarticletitle{Graph attention networks}.
\newblock \bibinfo{journal}{\emph{arXiv preprint arXiv:1710.10903}}
  (\bibinfo{year}{2017}).
\newblock


\bibitem[\protect\citeauthoryear{Wang, Dou, Chen, Chen, Liu, and Yu}{Wang
  et~al\mbox{.}}{2021a}]%
        {DBLP:conf/bigdataconf/WangDCCLY21}
\bibfield{author}{\bibinfo{person}{Chen Wang}, \bibinfo{person}{Yingtong Dou},
  \bibinfo{person}{Min Chen}, \bibinfo{person}{Jia Chen},
  \bibinfo{person}{Zhiwei Liu}, {and} \bibinfo{person}{Philip~S. Yu}.}
  \bibinfo{year}{2021}\natexlab{a}.
\newblock \showarticletitle{Deep Fraud Detection on Non-attributed Graph}. In
  \bibinfo{booktitle}{\emph{2021 {IEEE} International Conference on Big Data
  (Big Data), Orlando, FL, USA, December 15-18, 2021}}.
  \bibinfo{publisher}{{IEEE}}, \bibinfo{pages}{5470--5473}.
\newblock
\urldef\tempurl%
\url{https://doi.org/10.1109/BigData52589.2021.9672028}
\showDOI{\tempurl}


\bibitem[\protect\citeauthoryear{Wang, Zhang, Zhang, Leskovec, Zhao, Li, and
  Wang}{Wang et~al\mbox{.}}{2019c}]%
        {wang2019knowledge}
\bibfield{author}{\bibinfo{person}{Hongwei Wang}, \bibinfo{person}{Fuzheng
  Zhang}, \bibinfo{person}{Mengdi Zhang}, \bibinfo{person}{Jure Leskovec},
  \bibinfo{person}{Miao Zhao}, \bibinfo{person}{Wenjie Li}, {and}
  \bibinfo{person}{Zhongyuan Wang}.} \bibinfo{year}{2019}\natexlab{c}.
\newblock \showarticletitle{Knowledge-aware graph neural networks with label
  smoothness regularization for recommender systems}. In
  \bibinfo{booktitle}{\emph{Proceedings of the 25th ACM SIGKDD international
  conference on knowledge discovery \& data mining}}.
  \bibinfo{pages}{968--977}.
\newblock


\bibitem[\protect\citeauthoryear{Wang, He, Cao, Liu, and Chua}{Wang
  et~al\mbox{.}}{2019a}]%
        {wang2019kgat}
\bibfield{author}{\bibinfo{person}{Xiang Wang}, \bibinfo{person}{Xiangnan He},
  \bibinfo{person}{Yixin Cao}, \bibinfo{person}{Meng Liu}, {and}
  \bibinfo{person}{Tat-Seng Chua}.} \bibinfo{year}{2019}\natexlab{a}.
\newblock \showarticletitle{Kgat: Knowledge graph attention network for
  recommendation}. In \bibinfo{booktitle}{\emph{Proceedings of the 25th ACM
  SIGKDD International Conference on Knowledge Discovery \& Data Mining}}.
  \bibinfo{pages}{950--958}.
\newblock


\bibitem[\protect\citeauthoryear{Wang, He, Wang, Feng, and Chua}{Wang
  et~al\mbox{.}}{2019b}]%
        {wang2019neural}
\bibfield{author}{\bibinfo{person}{Xiang Wang}, \bibinfo{person}{Xiangnan He},
  \bibinfo{person}{Meng Wang}, \bibinfo{person}{Fuli Feng}, {and}
  \bibinfo{person}{Tat-Seng Chua}.} \bibinfo{year}{2019}\natexlab{b}.
\newblock \showarticletitle{Neural graph collaborative filtering}. In
  \bibinfo{booktitle}{\emph{Proceedings of the 42nd international ACM SIGIR
  conference on Research and development in Information Retrieval}}.
  \bibinfo{pages}{165--174}.
\newblock


\bibitem[\protect\citeauthoryear{Wang, Liu, Fan, Sun, and Yu}{Wang
  et~al\mbox{.}}{2021b}]%
        {wang2021dskreg}
\bibfield{author}{\bibinfo{person}{Yu Wang}, \bibinfo{person}{Zhiwei Liu},
  \bibinfo{person}{Ziwei Fan}, \bibinfo{person}{Lichao Sun}, {and}
  \bibinfo{person}{Philip~S Yu}.} \bibinfo{year}{2021}\natexlab{b}.
\newblock \showarticletitle{DSKReG: Differentiable Sampling on Knowledge Graph
  for Recommendation with Relational GNN}. In
  \bibinfo{booktitle}{\emph{Proceedings of the 30th ACM International
  Conference on Information \& Knowledge Management}}.
  \bibinfo{pages}{3513--3517}.
\newblock


\bibitem[\protect\citeauthoryear{Wang, Shen, and Cremers}{Wang
  et~al\mbox{.}}{2021c}]%
        {wang2021explicit}
\bibfield{author}{\bibinfo{person}{Yu Wang}, \bibinfo{person}{Yuesong Shen},
  {and} \bibinfo{person}{Daniel Cremers}.} \bibinfo{year}{2021}\natexlab{c}.
\newblock \showarticletitle{Explicit pairwise factorized graph neural network
  for semi-supervised node classification}. In
  \bibinfo{booktitle}{\emph{Uncertainty in Artificial Intelligence}}. PMLR,
  \bibinfo{pages}{1979--1987}.
\newblock


\bibitem[\protect\citeauthoryear{Wu, Sun, Fu, Hong, Wang, and Wang}{Wu
  et~al\mbox{.}}{2019a}]%
        {wu2019neural}
\bibfield{author}{\bibinfo{person}{Le Wu}, \bibinfo{person}{Peijie Sun},
  \bibinfo{person}{Yanjie Fu}, \bibinfo{person}{Richang Hong},
  \bibinfo{person}{Xiting Wang}, {and} \bibinfo{person}{Meng Wang}.}
  \bibinfo{year}{2019}\natexlab{a}.
\newblock \showarticletitle{A neural influence diffusion model for social
  recommendation}. In \bibinfo{booktitle}{\emph{Proceedings of the 42nd
  International {ACM} {SIGIR} Conference on Research and Development in
  Information Retrieval, {SIGIR} 2019, Paris, France, July 21-25, 2019}}.
  \bibinfo{publisher}{{ACM}}, \bibinfo{pages}{235--244}.
\newblock


\bibitem[\protect\citeauthoryear{Wu, Zhang, Gao, He, Weng, Gao, and Chen}{Wu
  et~al\mbox{.}}{2019b}]%
        {wu2019dual}
\bibfield{author}{\bibinfo{person}{Qitian Wu}, \bibinfo{person}{Hengrui Zhang},
  \bibinfo{person}{Xiaofeng Gao}, \bibinfo{person}{Peng He},
  \bibinfo{person}{Paul Weng}, \bibinfo{person}{Han Gao}, {and}
  \bibinfo{person}{Guihai Chen}.} \bibinfo{year}{2019}\natexlab{b}.
\newblock \showarticletitle{Dual graph attention networks for deep latent
  representation of multifaceted social effects in recommender systems}. In
  \bibinfo{booktitle}{\emph{The World Wide Web Conference}}.
  \bibinfo{pages}{2091--2102}.
\newblock


\bibitem[\protect\citeauthoryear{Wu, Pan, Chen, Long, Zhang, and Yu}{Wu
  et~al\mbox{.}}{2021}]%
        {DBLP:journals/tnn/WuPCLZY21}
\bibfield{author}{\bibinfo{person}{Zonghan Wu}, \bibinfo{person}{Shirui Pan},
  \bibinfo{person}{Fengwen Chen}, \bibinfo{person}{Guodong Long},
  \bibinfo{person}{Chengqi Zhang}, {and} \bibinfo{person}{Philip~S. Yu}.}
  \bibinfo{year}{2021}\natexlab{}.
\newblock \showarticletitle{A Comprehensive Survey on Graph Neural Networks}.
\newblock \bibinfo{journal}{\emph{{IEEE} Trans. Neural Networks Learn. Syst.}}
  \bibinfo{volume}{32}, \bibinfo{number}{1} (\bibinfo{year}{2021}),
  \bibinfo{pages}{4--24}.
\newblock
\urldef\tempurl%
\url{https://doi.org/10.1109/TNNLS.2020.2978386}
\showDOI{\tempurl}


\bibitem[\protect\citeauthoryear{Xu, Ruan, Korpeoglu, Kumar, and Achan}{Xu
  et~al\mbox{.}}{2020}]%
        {xu2020product}
\bibfield{author}{\bibinfo{person}{Da Xu}, \bibinfo{person}{Chuanwei Ruan},
  \bibinfo{person}{Evren Korpeoglu}, \bibinfo{person}{Sushant Kumar}, {and}
  \bibinfo{person}{Kannan Achan}.} \bibinfo{year}{2020}\natexlab{}.
\newblock \showarticletitle{Product knowledge graph embedding for e-commerce}.
  In \bibinfo{booktitle}{\emph{Proceedings of the 13th international conference
  on web search and data mining}}. \bibinfo{pages}{672--680}.
\newblock


\bibitem[\protect\citeauthoryear{Xu, Hu, Leskovec, and Jegelka}{Xu
  et~al\mbox{.}}{2019}]%
        {DBLP:conf/iclr/XuHLJ19}
\bibfield{author}{\bibinfo{person}{Keyulu Xu}, \bibinfo{person}{Weihua Hu},
  \bibinfo{person}{Jure Leskovec}, {and} \bibinfo{person}{Stefanie Jegelka}.}
  \bibinfo{year}{2019}\natexlab{}.
\newblock \showarticletitle{How Powerful are Graph Neural Networks?}. In
  \bibinfo{booktitle}{\emph{7th International Conference on Learning
  Representations, {ICLR} 2019, New Orleans, LA, USA, May 6-9, 2019}}.
  \bibinfo{publisher}{OpenReview.net}.
\newblock
\urldef\tempurl%
\url{https://openreview.net/forum?id=ryGs6iA5Km}
\showURL{%
\tempurl}


\bibitem[\protect\citeauthoryear{Yang, Liu, Dou, Ma, and Yu}{Yang
  et~al\mbox{.}}{2021}]%
        {yang21consisrec}
\bibfield{author}{\bibinfo{person}{Liangwei Yang}, \bibinfo{person}{Zhiwei
  Liu}, \bibinfo{person}{Yingtong Dou}, \bibinfo{person}{Jing Ma}, {and}
  \bibinfo{person}{Philip~S. Yu}.} \bibinfo{year}{2021}\natexlab{}.
\newblock \showarticletitle{ConsisRec: Enhancing {GNN} for Social
  Recommendation via Consistent Neighbor Aggregation}. In
  \bibinfo{booktitle}{\emph{{SIGIR} '21: The 44th International {ACM} {SIGIR}
  Conference on Research and Development in Information Retrieval, Virtual
  Event, Canada, July 11-15, 2021}},
  \bibfield{editor}{\bibinfo{person}{Fernando Diaz}, \bibinfo{person}{Chirag
  Shah}, \bibinfo{person}{Torsten Suel}, \bibinfo{person}{Pablo Castells},
  \bibinfo{person}{Rosie Jones}, {and} \bibinfo{person}{Tetsuya Sakai}} (Eds.).
  \bibinfo{publisher}{{ACM}}, \bibinfo{pages}{2141--2145}.
\newblock
\urldef\tempurl%
\url{https://doi.org/10.1145/3404835.3463028}
\showDOI{\tempurl}


\bibitem[\protect\citeauthoryear{Yi, Hong, Zhong, Liu, and Rajan}{Yi
  et~al\mbox{.}}{2014}]%
        {yi2014beyond}
\bibfield{author}{\bibinfo{person}{Xing Yi}, \bibinfo{person}{Liangjie Hong},
  \bibinfo{person}{Erheng Zhong}, \bibinfo{person}{Nanthan~Nan Liu}, {and}
  \bibinfo{person}{Suju Rajan}.} \bibinfo{year}{2014}\natexlab{}.
\newblock \showarticletitle{Beyond clicks: dwell time for personalization}. In
  \bibinfo{booktitle}{\emph{Proceedings of the 8th ACM Conference on
  Recommender systems}}. \bibinfo{pages}{113--120}.
\newblock


\bibitem[\protect\citeauthoryear{Ying, He, Chen, Eksombatchai, Hamilton, and
  Leskovec}{Ying et~al\mbox{.}}{2018}]%
        {ying2018graph}
\bibfield{author}{\bibinfo{person}{Rex Ying}, \bibinfo{person}{Ruining He},
  \bibinfo{person}{Kaifeng Chen}, \bibinfo{person}{Pong Eksombatchai},
  \bibinfo{person}{William~L Hamilton}, {and} \bibinfo{person}{Jure Leskovec}.}
  \bibinfo{year}{2018}\natexlab{}.
\newblock \showarticletitle{Graph convolutional neural networks for web-scale
  recommender systems}. In \bibinfo{booktitle}{\emph{Proceedings of the 24th
  ACM SIGKDD International Conference on Knowledge Discovery \& Data Mining}}.
  \bibinfo{pages}{974--983}.
\newblock


\bibitem[\protect\citeauthoryear{Yu, Gao, Li, Yin, and Liu}{Yu
  et~al\mbox{.}}{2018}]%
        {DBLP:conf/cikm/Yu0LYL18}
\bibfield{author}{\bibinfo{person}{Junliang Yu}, \bibinfo{person}{Min Gao},
  \bibinfo{person}{Jundong Li}, \bibinfo{person}{Hongzhi Yin}, {and}
  \bibinfo{person}{Huan Liu}.} \bibinfo{year}{2018}\natexlab{}.
\newblock \showarticletitle{Adaptive Implicit Friends Identification over
  Heterogeneous Network for Social Recommendation}. In
  \bibinfo{booktitle}{\emph{Proceedings of the 27th {ACM} International
  Conference on Information and Knowledge Management, {CIKM} 2018, Torino,
  Italy, October 22-26, 2018}}, \bibfield{editor}{\bibinfo{person}{Alfredo
  Cuzzocrea}, \bibinfo{person}{James Allan}, \bibinfo{person}{Norman~W. Paton},
  \bibinfo{person}{Divesh Srivastava}, \bibinfo{person}{Rakesh Agrawal},
  \bibinfo{person}{Andrei~Z. Broder}, \bibinfo{person}{Mohammed~J. Zaki},
  \bibinfo{person}{K.~Sel{\c{c}}uk Candan}, \bibinfo{person}{Alexandros
  Labrinidis}, \bibinfo{person}{Assaf Schuster}, {and} \bibinfo{person}{Haixun
  Wang}} (Eds.). \bibinfo{publisher}{{ACM}}, \bibinfo{pages}{357--366}.
\newblock


\bibitem[\protect\citeauthoryear{Zhang and McAuley}{Zhang and McAuley}{2020}]%
        {DBLP:conf/sdm/ZhangM20}
\bibfield{author}{\bibinfo{person}{Hengrui Zhang} {and}
  \bibinfo{person}{Julian~J. McAuley}.} \bibinfo{year}{2020}\natexlab{}.
\newblock \showarticletitle{Stacked Mixed-Order Graph Convolutional Networks
  for Collaborative Filtering}. In \bibinfo{booktitle}{\emph{Proceedings of the
  2020 {SIAM} International Conference on Data Mining, {SDM} 2020, Cincinnati,
  Ohio, USA, May 7-9, 2020}}, \bibfield{editor}{\bibinfo{person}{Carlotta
  Demeniconi} {and} \bibinfo{person}{Nitesh~V. Chawla}} (Eds.).
  \bibinfo{publisher}{{SIAM}}, \bibinfo{pages}{73--81}.
\newblock
\urldef\tempurl%
\url{https://doi.org/10.1137/1.9781611976236.9}
\showDOI{\tempurl}


\bibitem[\protect\citeauthoryear{Zhang, Wu, Yan, Wipf, and Yu}{Zhang
  et~al\mbox{.}}{2021}]%
        {DBLP:journals/corr/abs-2106-12484}
\bibfield{author}{\bibinfo{person}{Hengrui Zhang}, \bibinfo{person}{Qitian Wu},
  \bibinfo{person}{Junchi Yan}, \bibinfo{person}{David Wipf}, {and}
  \bibinfo{person}{Philip~S. Yu}.} \bibinfo{year}{2021}\natexlab{}.
\newblock \showarticletitle{From Canonical Correlation Analysis to
  Self-supervised Graph Neural Networks}.
\newblock \bibinfo{journal}{\emph{CoRR}}  \bibinfo{volume}{abs/2106.12484}
  (\bibinfo{year}{2021}).
\newblock
\showeprint[arXiv]{2106.12484}
\urldef\tempurl%
\url{https://arxiv.org/abs/2106.12484}
\showURL{%
\tempurl}


\bibitem[\protect\citeauthoryear{Zheng, Fan, Lu, Zhang, and Yu}{Zheng
  et~al\mbox{.}}{2019}]%
        {zheng2019gated}
\bibfield{author}{\bibinfo{person}{Lei Zheng}, \bibinfo{person}{Ziwei Fan},
  \bibinfo{person}{Chun-Ta Lu}, \bibinfo{person}{Jiawei Zhang}, {and}
  \bibinfo{person}{Philip~S Yu}.} \bibinfo{year}{2019}\natexlab{}.
\newblock \showarticletitle{Gated Spectral Units: Modeling Co-evolving Patterns
  for Sequential Recommendation}. In \bibinfo{booktitle}{\emph{Proceedings of
  the 42nd International ACM SIGIR Conference on Research and Development in
  Information Retrieval}}. \bibinfo{pages}{1077--1080}.
\newblock


\bibitem[\protect\citeauthoryear{Zhou, Xu, Wu, Taghavi, Korpeoglu, Achan, and
  He}{Zhou et~al\mbox{.}}{2021}]%
        {zhou2021pure}
\bibfield{author}{\bibinfo{person}{Yao Zhou}, \bibinfo{person}{Jianpeng Xu},
  \bibinfo{person}{Jun Wu}, \bibinfo{person}{Zeinab Taghavi},
  \bibinfo{person}{Evren Korpeoglu}, \bibinfo{person}{Kannan Achan}, {and}
  \bibinfo{person}{Jingrui He}.} \bibinfo{year}{2021}\natexlab{}.
\newblock \showarticletitle{PURE: Positive-Unlabeled Recommendation with
  Generative Adversarial Network}. In \bibinfo{booktitle}{\emph{Proceedings of
  the 27th ACM SIGKDD Conference on Knowledge Discovery \& Data Mining}}.
  \bibinfo{pages}{2409--2419}.
\newblock


\end{thebibliography}

\newpage

\appendix

\section{Data Analysis}
In this section, we present more details regarding the data analyses. 

\begin{table}[htbp]
\caption{Number of players w.r.t. different number of game interactions}
\label{tab:user_num_interaction}
\begin{tabular}{c|c}
\toprule
\textbf{Interacted game number}&\textbf{Player number} \\
\hline
1 & 10,764,869 \\
2 & 5,113,208 \\
3 & 2,960,352 \\
4 & 3,847,987 \\
5 & 1,483,215 \\
6 & 2,773,800 \\
7 & 1,165,742 \\
8 & 1,743,871 \\
9 & 749,208 \\
$\geq$ 10 & 8,170,542\\
\bottomrule
\end{tabular} 
\end{table}

Table~\ref{tab:user_num_interaction} shows the number of players with respect to different number of game interactions. As discussed in Section~\ref{sec:Personalization}, a large portion of players only interact with limited number of games. The data sparsity causes challenges when recommending for these cold-start players. In SCGRec, we aim to mitigate this problem by players' social graph and games' context graph. Players' social graph can link players to more games through social friends, and games' context graph can learn the relevance information among games.

\begin{table}[htbp]
\caption{Number of players w.r.t. different number of friends}
\label{tab:social_num}
\begin{tabular}{c|c}
\toprule
\textbf{Number of Social Friends}&\textbf{Number of players} \\
\hline
1 & 9,025,326 \\
2 & 4,286,860 \\
3 & 2,769,634 \\
4 & 2,012,605 \\
5 & 1,564,368 \\
6 & 1,269,972 \\
7 & 1,056,377 \\
8 & 895,596 \\
9 & 770,913 \\
$\geq$ 10 & 9,468,130\\
\bottomrule
\end{tabular} 
\end{table}

\begin{figure}[htbp]
         \centering
         \includegraphics[width=0.3\textwidth]{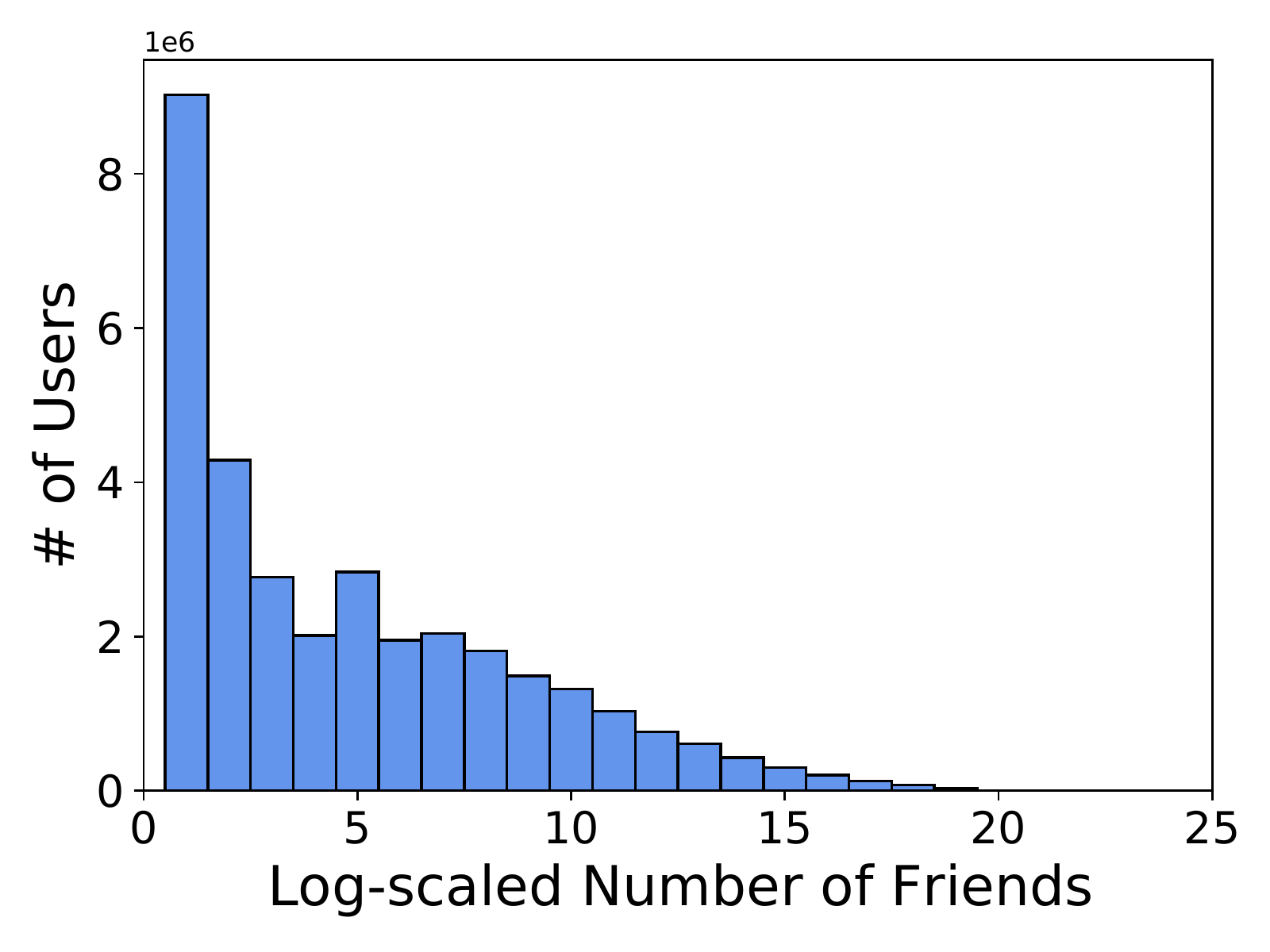}
         \caption{Players' number of friends distribution}
         \label{fig:friendship_num}
\end{figure}

As shown in Figure~\ref{fig:friendship_num}, players number of friends follows power-law distribution. Table~\ref{tab:social_num} shows the exact number of players with respect to different number of friends. It can be observed that a large portion of players have limited number of friends. As side information, social graph can link more games through social neighbor, which is more important for cold-start players.

\begin{table}[htbp]
\caption{Pearson correlation coefficient w.r.t. genres}
\label{tab:data_genre}
\begin{tabular}{l|c|c}
\toprule
\textbf{Genre}&\textbf{ player-friend} & \textbf{player-random}\\
\hline
Action & 0.4597 & 0.1782\\
Adventure & 0.1449 & 0.0546\\
RPG & 0.2201 & 0.0661\\
Simulation & 0.3722 & 0.0828\\
Strategy & 0.3552 & 0.1070\\
\bottomrule
\end{tabular} 
\end{table}

Table~\ref{tab:data_genre} shows pearson correlation coefficient between user-friend and user-random. Similar to Table~\ref{tab:Pearson_Correlation} in Section~\ref{sec:social_connection}, we can have two observations. Firstly, the pearson correlation coefficient between player and their friends is much higher than random players. It shows friends tend to have preference on the same game genre. Secondly, the social influence on different genres are different. For example, the pearson correlation coefficient of action genre is much larger than adventure genre. It shows social influence has a different level of impact on different genres.


\begin{figure}[htbp]
         \centering
         \includegraphics[width=0.45\textwidth]{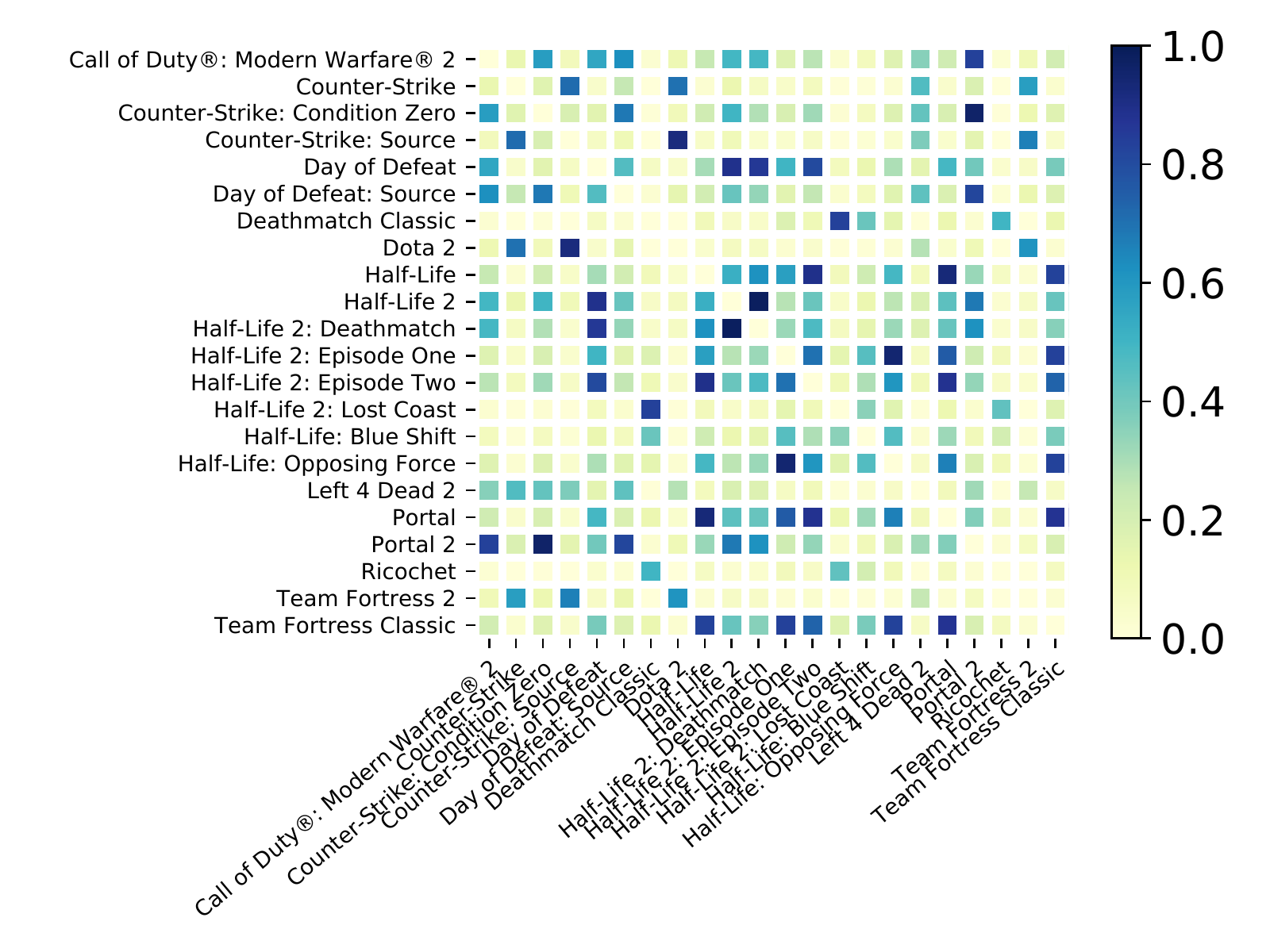}
         \caption{Game-Game Co-Dwelling}
         \label{fig:Co-dwelling_relation}
\end{figure}

Figure~\ref{fig:Co-dwelling_relation} shows co-dwelling relation among part of the games. Similar to co-purchase relation shown in Figure~\ref{fig:Co-Purchase_relation}, figure~\ref{fig:Co-dwelling_relation} also shows a high correlation between game pairs with respect to co-dwelling time. For example, co-dwelling time of game "Half-Life" and "Half-Life 2: Episode Two" is very long. It shows an apparent signal of game co-dwelling context.

\section{Experimental Setting}
Besides the data analyses, we also include more experimental settings. 
\begin{table}[!h]
\caption{Best Hyper-parameter Setting}
\label{tab:hyper-param}
\begin{tabular}{l|c|c}
\toprule
\textbf{Hyper-parameter}& \textbf{Search range} & \textbf{Best setting}\\
\hline
Embedding size & \{4,8,16,32,64\} & 32 \\
Learning rate & \{0.03,0.01,0.001\} & 0.03 \\
Batch size & \{128,256,512,1024\} & 1024 \\
$\lambda$ & \{1e-3, 1e-4, 1e-5\} & 1e-4 \\
$W_{social}$ & \{0.0,0.1,0.2,0.3,0.4,0.5\} & 0.1 \\
$W_{context}$ & \{0.0,0.1,0.2,0.3,0.4,0.5\} & 0.5 \\
\bottomrule
\end{tabular} 
\end{table}

The search range and best setting of hyper-parameters are given in Table~\ref{tab:hyper-param}.$W_{self}$ is calculated by $W_{self}=1-W_{social}-W_{context}$. The same search range is applied to all the baselines.

Table~\ref{tab:baseline comparison} presents comparisons among all methods. 
Popularity (time) and Popularity (count) are non-personalized recommendation models. They directly recommend games based on popularity. 
LightGCN, GIN, PinSAGE and GAT utilize only user-game interaction data, and learn the node embedding by different GNN models. 
RGCN and our proposed SCGRec incorporate all the information including user-game interaction, social graph and game contextualization. A brief description of all the baselines are given as follows:

\begin{table}
\caption{Model comparison}
\label{tab:baseline comparison}
\begin{tabular}{l|c|c|c}
\toprule
\textbf{Model}& \textbf{Social} & \textbf{Context} & \textbf{Personalization}\\
\hline
Popularity (time) & $\times$ & $\times$ & $\times$\\
Popularity (count) & $\times$ & $\times$ & $\times$\\
LightGCN & $\times$& $\times$ & \checkmark\\
GIN  &$\times$ & $\times$ & \checkmark\\
PinSAGE  &$\times$ &$\times$ & \checkmark\\
GAT  & $\times$& $\times$ & \checkmark\\
RGCN  & \checkmark&\checkmark & \checkmark\\
SCGRec  &\checkmark &\checkmark & \checkmark\\

\bottomrule
\end{tabular} 
\end{table}



\begin{itemize}[leftmargin=*]
    \item Popularity (time): Same as Popularity (count), except that we measure the game popularity based on total dwelling time.
    \item Popularity (count): Directly rank all the games based on the number of players. Each player is recommended with the same game list with no personalization.
    \item LightGCN~\cite{he2020lightgcn}: LightGCN simplifies Graph Convolution Network (GCN)~\cite{DBLP:conf/iclr/KipfW17} by removing transformation matrix, non-linear transformation, and self-loop, which is the SOTA GNN recommendation model. 
    \item RGCN~\cite{schlichtkrull2018modeling}: Relational Graph Convolution Network learns node embedding from all the side information. Each relation type is assigned one GCN layer for relation-specific aggregation.
    \item GIN~\cite{DBLP:conf/iclr/XuHLJ19}: Graph Isomorphism Network is a simple graph neural network that expects to achieve the ability as the Weisfeiler-Lehman graph isomorphism test.
    \item PinSAGE~\cite{ying2018graph}: PinSAGE combines efficient random walks in graph convolution to learn node embedding on large-scale recommendation bipartite graphs.
    \item GAT~\cite{velivckovic2017graph}: Graph Attention Network firstly learns the attention scores for neighbors before its neighboring aggregation.
\end{itemize}

\end{document}